\documentclass[runningheads]{llncs}
\usepackage{calc}
\usepackage{tikz}
\usetikzlibrary{decorations.pathreplacing}
\usepackage{pgfplots}
\usepgfplotslibrary{polar}
\pgfplotsset{compat=newest}
\usetikzlibrary{decorations.markings}
\tikzstyle{vertex}=[circle, draw, inner sep=0pt, minimum size=6pt]

\tikzstyle{vertex}=[circle, draw, inner sep=0pt, minimum size=3pt]
\tikzstyle{triangle}=[draw, shape=regular polygon, regular polygon sides=3,draw,inner sep=0pt,minimum
size=0.3cm, anchor=apex]
\usepackage{picture,xcolor}
\usepackage{graphicx}
\usepackage{amsmath}
\usepackage{amssymb}
\let\proof\relax
 
\usepackage{amsthm}
\usepackage{amsfonts}
\usepackage{xcolor}
\usepackage{graphicx}
\usepackage{tikz}
\usetikzlibrary{circuits.logic.US}
\usetikzlibrary{positioning}
\usetikzlibrary{matrix}
\usepackage{graphicx}
\newcommand{\boundellipse}[3]
{(#1) ellipse (#2 and #3)
}
\definecolor{magenta}{rgb}{0.8, 0.0, 0.8}
\definecolor{cyan}{rgb}{0.0, 1.0, 1.0}
\definecolor{blue1}{rgb}{0.1, 0.6, 0.01}
\definecolor{blue}{rgb}{0.10, 0.50, 1}
\definecolor{brown}{rgb}{0.65, 0.16, 0.16}
\definecolor{magenta}{rgb}{0.8, 0.0, 0.8}
\definecolor{cyan}{rgb}{0.0, 1.0, 1.0}
\definecolor{green1}{rgb}{0.1, 0.6, 0.01}
\definecolor{green}{rgb}{0, 1, 0}
\definecolor{brown}{rgb}{0.65, 0.16, 0.16}
\usepackage{graphicx}
\usepackage{amsmath}
\usepackage{amsfonts}
\usepackage{xcolor}
\usepackage{amssymb}
\usepackage{graphicx}
\usepackage{tikz}
\usetikzlibrary{circuits.logic.US}
\usetikzlibrary{positioning}
\usetikzlibrary{matrix}
\usepackage{graphicx}
\usepackage{amsmath}
\definecolor{magenta}{rgb}{0.8, 0.0, 0.8}
\definecolor{cyan}{rgb}{0.0, 1.0, 1.0}
\definecolor{green1}{rgb}{0.1, 0.6, 0.01}
\definecolor{green}{rgb}{0.11, 0.35, 0.02}
\definecolor{brown}{rgb}{0.65, 0.16, 0.16}
\definecolor{battleshipgrey}{rgb}{0.52, 0.52, 0.51}
\definecolor{babyblue}{rgb}{0.54, 0.81, 0.94}
\definecolor{brightgreen}{rgb}{0.4, 1.0, 0.0}
\definecolor{dimgray}{rgb}{0.41, 0.41, 0.41}

\begin{document}

\title{The Harmless Set Problem}
\titlerunning{The Harmless Set Problem}
%
%
%
%
\author{Ajinkya Gaikwad\thanks{The first author  gratefully acknowledges support from the Ministry of Human Resource Development, 
 Government of India, under Prime Minister's Research Fellowship Scheme (No. MRF-192002-211). } \and Soumen Maity\thanks{The 
 second author's research was supported in part by the Science and Engineering Research Board (SERB), Govt. of India, under Sanction Order No.
MTR/2018/001025.}}
\authorrunning{A.\,Gaikwad and S.\,Maity}
%
\institute{Indian Institute of Science Education and Research, Pune, India 
\email{\texttt{ajinkya.gaikwad@students.iiserpune.ac.in;}}
\email{\texttt{soumen@iiserpune.ac.in}}
}

\maketitle              
\begin{abstract}

Given a graph $G = (V,E)$, a threshold function $t~ :~ V \rightarrow \mathbb{N}$  and an integer $k$,
we study the {\sc Harmless Set} problem, where the goal is to find a subset of vertices 
$S \subseteq V$ of size at least $k$ such that every vertex $v\in V$ has less than $t(v)$ neighbors in $S$. We enhance our understanding of the problem from the 
viewpoint of parameterized complexity. Our focus lies on parameters that measure the structural properties of the input instance. We  show that the problem is W[1]-hard parameterized by a wide range of fairly restrictive structural parameters such as the feedback vertex set number, pathwidth, treedepth, and even the size of a minimum vertex deletion set into graphs of pathwidth and treedepth at most three. On dense graphs, we show that the  problem is W[1]-hard parameterized by cluster vertex deletion number. We also show that the  {\sc Harmless Set} problem with majority thresholds is W[1]-hard when parameterized by the treewidth of the input graph. We prove that the {\sc Harmless Set} problem  can be solved in polynomial time on graph with bounded cliquewidth. On the positive side, we obtain fixed-parameter algorithms for the problem with respect to  neighbourhood diversity, twin  cover and vertex integrity of the input graph. We show that the problem 
parameterized by the solution size  is fixed parameter tractable on planar graphs. We thereby resolve two open questions stated in C. Bazgan and  M. Chopin (2014) concerning the complexity of {\sc Harmless Set} parameterized by the treewidth of the input graph and
on planar graphs
with respect to the solution size. 
\end{abstract}

\section{Introduction} 
Social networks are used  not only to stay in touch with friends and family, but also to spread and receive information on specific products and services.
The spread of information through social networks is a well-documented and well-studied 
topic.  Kempe, Kleinberg, and Tardos \cite{KempeTardos} initiated a model to study the spread of influence through a social network. One of the most well known problems that appear in this context is {\sc Target Set Selection} introduced by Chen \cite{ChenSIAM} and defined as follows. We are given a graph, modeling a social network, where each node $v$ has a (fixed) threshold $t(v)$, the node will adopt a new product if $t(v)$ of its neighbors adopt it. Our goal is to find a small set $S$ of nodes such that targeting the product to $S$ would lead to adoption of the product by a large number of nodes in the graph.
This problem may occur for example in the context of 
disease propagation, viral marketing or even faults in 
distributed computing \cite{DREYER20091615,PELEG}. This problem received considerable
attention in a series of papers from classical complexity \cite{DREYER20091615,CENTENO20113693,Chiang,JGAA-244}, 
polynomial time approximability \cite{ChenSIAM,Aazami}, parameterized 
approximability \cite{BazganCocoon}, and parameterized complexity \cite{BENZWI201187,ChopinTCS,Nichterlein}. A natural research direction considering this fact is to look for the complexity of variants or constrained version of this problem. 
Bazgan and Chopin \cite{Bazgan-DO} followed this line of research and introduced the notion of harmless set. Throughout this article, $G=(V,E)$ denotes a finite, simple and undirected graph. We denote by $V(G)$ and $E(G)$ its vertex and edge set respectively.
For a vertex $v\in V$, we use $N(v)=\{u~:~(u,v)\in E(G)\}$ to denote the (open) neighbourhood 
of  $v$ in $G$. The degree $d(v)$ of a vertex 
$v\in V(G)$ is $|N(v)|$. For a subset $S\subseteq V(G)$, we use $N_S(v)=\{u \in S~:~(u,v)\in E(G)\}$ to denote the (open) neighbourhood 
of vertex $v$ in $S$. The degree $d_S(v)$ of a vertex $v\in V(G)$ in $S$ 
is $|N_S(v)|$. A harmless set consists of a set $S$ of vertices with the property that no propagation occurs if any subset of $S$ gets activated. In other words, a harmless set is defined as a converse notion  of a target set. More formally, 
\begin{definition}\rm \cite{Bazgan-DO}
A set $S\subseteq V$ is a \emph{harmless set} of $G=(V,E)$, if every vertex  $v\in V$ has less
than $t(v)$ neighbours in $S$. 
\end{definition} 
\noindent Note that in the definition of harmless set, the threshold condition is 
imposed on every vertex, including those in the solution $S$. As mentioned in \cite{Bazgan-DO}, another perhaps more natural definition could have been a set 
$S$ such that every vertex $v\notin S$ has less than $t(v)$ neighbours in $S$.
This definition creates two problems. First, it makes {\sc Harmless Set} problem 
meaningless as the whole set of vertices of the input graph would be a trivial solution. 
Second, there might be some propagation steps inside $S$ if some vertices are activated in $S$. 
In this paper, we consider the {\sc Harmless Set} problem
  under structural parameters. We define the problem as follows:
\vspace{3mm}
    \\
 \noindent   \fbox
    {\begin{minipage}{33.7em}\label{FFVS }
       {\sc Harmless Set}\\
        \noindent{\bf Input:} A graph $G=(V,E)$, a threshold function $t :V \rightarrow \mathbb{N} $ where $1\leq t(v)\leq d(v)$ for every  $v \in V$, and an integer $k$.
    
        \noindent{\bf Question:} Is there a harmless set $S \subseteq V$ of size at 
        least $k$?
    \end{minipage} }
    \vspace{3mm}
    \\
The majority threshold is $t (v) = \lceil\frac{d(v)}{2}\rceil$  for all $v \in V$.  
We now review the concept of a tree decomposition, introduced by Robertson and Seymour in \cite{Neil}.
Treewidth is a  measure of how “tree-like” the graph is.
\begin{definition}\rm \cite{Downey} A {\it tree decomposition} of a graph $G=(V,E)$  is a tree $T$ together with a 
collection of subsets $X_t$ (called \emph{bags}) of $V$ labeled by the nodes $t$ of $T$ such that 
$\bigcup_{t\in T}X_t=V $ and (1) and (2) below hold:
\begin{enumerate}
			\item For every edge $uv \in E(G)$, there  is some $t$ such that $\{u,v\}\subseteq X_t$.
			\item  (Interpolation Property) If $t$ is a node on the unique path in $T$ from $t_1$ to $t_2$, then 
			$X_{t_1}\cap X_{t_2}\subseteq X_t$.
		\end{enumerate}
	\end{definition}
	
\begin{definition}\rm \cite{Downey} The {\it width} of a tree decomposition is
the maximum value of $|X_t|-1 $ taken over all the nodes $t$ of the tree $T$ of the decomposition.
The \emph{treewidth} $tw(G)$ of a graph $G$  is the  minimum width among all possible tree decompositions of $G$.
\end{definition} 
\begin{figure}
     \centering
 \begin{tikzpicture}[scale=1]
 \node[fill=black, circle, draw=black, inner sep=0, minimum size=0.15cm](b) at (0,3) [label=left: $b$] {};
 \node[fill=black, circle, draw=black, inner sep=0, minimum size=0.15cm](a) at (0,2) [label=left:$a$] {}; 
 \node[fill=black, circle, draw=black, inner sep=0, minimum size=0.15cm](c) at (2,2) [label=left: $c$] {};
 \node[fill=black, circle, draw=black, inner sep=0, minimum size=0.15cm](d) at (2,3) [label=above:$d$] {}; 
 \node[fill=black, circle, draw=black, inner sep=0, minimum size=0.15cm](h) at (3,4) [label=above:$h$] {}; 
 \node[fill=black, circle, draw=black, inner sep=0, minimum size=0.15cm](e) at (4,3) [label=above:$e$] {}; 
 \node[fill=black, circle, draw=black, inner sep=0, minimum size=0.15cm](f) at (2,1) [label=left:$f$] {}; 
 \node[fill=black, circle, draw=black, inner sep=0, minimum size=0.15cm](g) at (3,2) [label=right:$g$] {}; 
 \node[thick, circle,draw, minimum size=.1cm ] (bd) at (7,5) []{{$b,d$}};
\node[thick, circle,draw, minimum size=.1cm ] (abd) at (6,4) []{{$a,b,d$}};
\node[thick, circle,draw, minimum size=.1cm ] (cbd) at (8,4) []{{$c,b,d$}};
\node[thick, circle,draw, minimum size=.1cm ] (cdh) at (9,3) []{{$c,d,h$}};
\node[thick, circle,draw, minimum size=.1cm ] (chg) at (8,2) []{{$c,h,g$}};
\node[thick, circle,draw, minimum size=.1cm ] (he) at (10,2) []{{$h,e$}};
\node[thick, circle,draw, minimum size=.1cm ] (cgf) at (7,1) []{{$c,g,f$}};
\path 
(bd) edge (abd)
(bd) edge (cbd)
(cbd) edge (cdh)
(cdh) edge (chg)
(cdh) edge (he)
(chg) edge (cgf);
\path
(a) edge (b)
(b) edge (d)
(b) edge (c)
(a) edge (d)
(d) edge (c)
(d) edge (h)
(h) edge (e)
(h) edge (c)
(c) edge (g)
(f) edge (g)
(h) edge (g)
(c) edge (f);

\end{tikzpicture}

     \caption{Example of a tree decomposition of width 2 }
     \label{fig:treewidth}
 \end{figure}
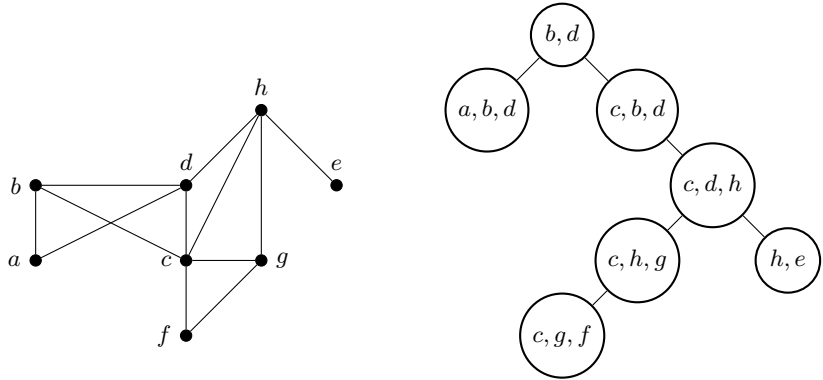

\begin{example}
Figure \ref{fig:treewidth} gives an example of a tree decomposition of width 2. 
\end{example}

\noindent A rooted forest is a disjoint union of rooted trees. Given a rooted forest $F$, its \emph{transitive closure} is a graph $H$ in which $V(H)$ contains all the nodes of the rooted forest, and $E(H)$ contain an edge between two vertices only if those two vertices form an ancestor-descendant pair in the forest $F$.

   \begin{definition}
        {\rm  The {\it treedepth} of a graph $G$ is the minimum height of a rooted forest $F$ whose transitive closure contains the graph $G$. It is denoted by $td(G)$.}
    \end{definition}

\begin{definition}\rm 
A set $S \subseteq V(G)$ is a \emph{vertex cover} of $G=(V,E)$ if  each edge in $E$ has at least one  endpoint in $S$.   The \emph{size} of a smallest vertex cover of  $G$ is the \emph{vertex cover number} of $G$.
\end{definition}

\noindent We recall a natural way of generalizing vertex cover to dense graphs. We relax the definition of vertex cover so that not all edges need to be covered.

\begin{definition}\rm 
An edge is a \emph{twin edge} if its incident vertices have the same neighborhood 
(excluding each other).
\end{definition}
\begin{definition}\rm 
A set $X\subseteq V(G)$ is a \emph{twin-cover} of $G$ if every edge in $G$ is either twin or incident to a vertex in $X$. We then say that $G$ has \emph{twin-cover number} $k$ if $k$ is the minimum possible size
of a twin-cover of $G$.
\end{definition}

\begin{definition}\rm 
A set $X\subseteq V(G)$ is a \emph{cluster vertex deletion} set of $G$ if $G \setminus X$ is a union of cliques.
\end{definition}

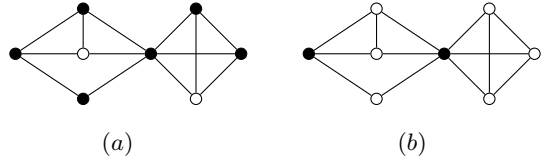
\begin{figure}[ht]
     \centering
 \begin{tikzpicture}[scale=0.6]
 \node[fill=black, circle, draw=black, inner sep=0, minimum size=0.15cm](x1) at (0,0) [label=left:] {};
 \node[fill=black, circle, draw=black, inner sep=0, minimum size=0.15cm](x2) at (-3,0) [label=left:] {};
  \node[circle, draw=black, inner sep=0, minimum size=0.15cm](x3) at (-1.5,0) [label=left:] {};
 \node[fill=black, circle, draw=black, inner sep=0, minimum size=0.15cm](x4) at (-1.5,1) [label=left:] {}; 
 \node[fill=black, circle, draw=black, inner sep=0, minimum size=0.15cm](x5) at (-1.5,-1) [label=left:] {};  
\node[fill=black, circle, draw=black, inner sep=0, minimum size=0.15cm](x6) at (2,0) [label=left:] {};
\node[fill=black, circle, draw=black, inner sep=0, minimum size=0.15cm](x7) at (1,1) [label=left:] {};  
\node[circle, draw=black, inner sep=0, minimum size=0.15cm](x8) at (1,-1) [label=left:] {}; 

\draw(x1)--(x3);
\draw(x3)--(x2);
\draw(x3)--(x4);
\draw(x1)--(x4);
\draw(x1)--(x5);
\draw(x2)--(x5);
\draw(x2)--(x4);
\draw(x1)--(x6);
\draw(x1)--(x7);
\draw(x1)--(x8);
\draw(x7)--(x8);
\draw(x7)--(x6);
\draw(x8)--(x6);

 \node(a) at (0,-2) [label=left:$(a)$] {};

 \node[fill=black, circle, draw=black, inner sep=0, minimum size=0.15cm](y1) at (6.5,0) [label=left:] {};
 \node[fill=black, circle, draw=black, inner sep=0, minimum size=0.15cm](y2) at (3.5,0) [label=left:] {};
  \node[circle, draw=black, inner sep=0, minimum size=0.15cm](y3) at (5,0) [label=left:] {};
 \node[ circle, draw=black, inner sep=0, minimum size=0.15cm](y4) at (5,1) [label=left:] {}; 
 \node[ circle, draw=black, inner sep=0, minimum size=0.15cm](y5) at (5,-1) [label=left:] {};  
\node[ circle, draw=black, inner sep=0, minimum size=0.15cm](y6) at (8.5,0) [label=left:] {};
\node[ circle, draw=black, inner sep=0, minimum size=0.15cm](y7) at (7.5,1) [label=left:] {};  
\node[circle, draw=black, inner sep=0, minimum size=0.15cm](y8) at (7.5,-1) [label=left:] {}; 

\draw(y1)--(y3);
\draw(y3)--(y2);
\draw(y3)--(y4);
\draw(y1)--(y4);
\draw(y1)--(y5);
\draw(y2)--(y5);
\draw(y2)--(y4);
\draw(y1)--(y6);
\draw(y1)--(y7);
\draw(y1)--(y8);
\draw(y7)--(y8);
\draw(y7)--(y6);
\draw(y8)--(y6);

 \node(a) at (6.5,-2) [label=left:$(b)$] {};

\end{tikzpicture}

     \caption{(a) A minimum size vertex cover (b) a minimum size twin cover of an example graph.}
     \label{twvc}
 \end{figure}

\noindent An illustration and comparison is provided in Figure \ref{twvc}. 

\vspace{3mm}
\begin{figure}
     \centering
 \begin{tikzpicture}[%
    auto,
    block/.style={
      rectangle,
      draw=black,
      thick,
      text width=3em,
      align=center,
      rounded corners,
      minimum height=1.5em
    }
]
    \draw (0,0) node[block] (vc) {\color{blue} vc}
          (2,1.5) node[block] (nd) {\color{blue} nd}
          (4,1.5) node[block] (tc) {\color{blue} tc}
          (-3,1.5) node[block] (vi) {\color{blue} vi}
          (-3,2.5) node[block] (td) { \color{red}  td}
          (-2,3.5) node[block] (fvs) {\color{red} fvs}
          (-4,3.5) node[block] (pw) { \color{red} pw}
          (2,3) node[block] (mw) {mw}
          (4,3) node[block] (cvd) { \color{red} cvd}
          (-3,4.5) node[block] (tw) {\color{red} tw}
          (-1,5.5) node[block] (cw) {\color{red} cw};
          
\draw[->, thick] (vc)--(vi); 
\draw[->, thick] (vc)--(nd); 
\draw[->, thick] (vc)--(tc); 
\draw[->, thick] (vi)--(td); 
\draw[->, thick] (td)--(pw); 
\draw[->, thick] (td)--(fvs); 
\draw[->, thick] (nd)--(mw); 
\draw[->, thick] (tc)--(mw); 
\draw[->, thick] (tc)--(cvd); 
\draw[->, thick] (pw)--(tw); 
\draw[->, thick] (fvs)--(tw); 
\draw[->, thick] (tw)--(cw); 
\draw[->, thick] (mw)--(cw); 
\draw[->, thick] (cvd)--(cw);




  \end{tikzpicture}

\caption{ Relationship between vertex cover (vc), neighbourhood diversity (nd), twin cover (tc), modular width (mw), cluster vertex deletion number (cvd), feedback vertex set (fvs), pathwidth (pw), treewidth (tw) and clique width (cw). 
Note that $A\rightarrow B$ means that there exists a function $f$ such that for 
all graphs, $f(A(G))\geq B(G)$. It also gives an overview of the parameterized complexity landscape for the {\sc Harmless Set} problem with general thresholds.
The problem is FPT parameterized by blue colored parameters and W[1]-hard 
when parameterized by red colored parameters. The problem remains unsettled when parameterized by mw. }

     \label{fig:parameters}
 \end{figure}
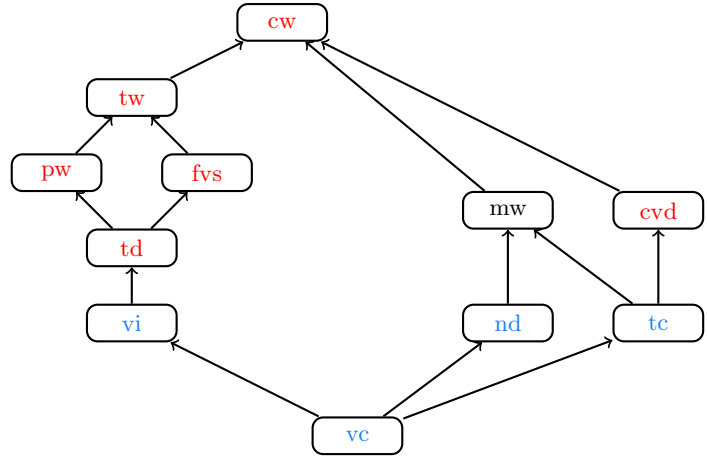
 

\noindent For the standard concepts in parameterized complexity, see the recent textbook by Cygan et al. \cite{marekcygan}.
\subsection{Our Results:}
 Our main results are as follows:     
\begin{itemize}
\item the {\sc Harmless Set} problem with general thresholds is FPT when parameterized by the neighbourhood diversity.

\item  the {\sc Harmless Set} problem with general thresholds is FPT when parameterized by twin cover of the input graph.

\item  the {\sc Harmless Set} problem with general thresholds is FPT when parameterized by vertex integrity of the input graph.

\item  the {\sc Harmless Set} problem with majority thresholds is W[1]-hard when parameterized by the treewidth of the graph.

 \item the {\sc Harmless Set} problem with general thresholds is W[1]-hard when 
 parameterized by the size of a vertex deletion set into
trees of height at most 3, even when restricted to bipartite graphs.

\item  the {\sc Harmless Set} problem with general thresholds is W[1]-hard when parameterized by cluster vertex deletion set.

\item the {\sc Harmless Set} problem with general thresholds can be solved in polynomial time on graph with bounded cliquewidth. 

\item the {\sc Harmless Set} problem with general thresholds is FPT when parameterized by the solution size when restricted to planar graphs.

\end{itemize}

\subsection{Known Results:}
Bazgan and Chopin \cite{Bazgan-DO} studied the parameterized complexity of {\sc Harmless Set} and the approximation of the associated maximization problem. When the parameter is $k$, they proved that the {\sc Harmless Set} problem is W[2]-complete in general and W[1]-complete if all thresholds are bounded by a constant. 
When each threshold is equal to the degree of the vertex, they showed that {\sc Harmless Set} is fixed-parameter tractable for parameter $k$ and the maximization version is APX-complete. 
They gave a polynomial-time algorithm for graphs of bounded treewidth and a polynomial-time approximation scheme for planar graphs. The parametric dual problem $(n - k)$-{\sc Harmless Set}  asks for the existence of a harmless set of size at least $n-k$. The parameter is $k$ and $n$ denotes the number of vertices in the input graph. They showed that the parametric dual problem $(n - k)$-{\sc Harmless Set} is fixed-parameter tractable for a large family of threshold functions.


\section{FPT algorithm parameterized by neighbourhood diversity}
 In this section, we present an FPT algorithm for the {\sc Harmless Set}  problem parameterized by neighbourhood diversity. 
  We say that two (distinct) vertices $u$ and $v$
 have the same {\it neighborhood type} if they share their respective neighborhoods, 
 that is, when $N(u)\setminus \{v\}=N(v)\setminus \{u\}$.
 If this is so we say that $u$
 and $v$ are {\it twins}. It is possible to distinguish true-twins (those joined by an edge) 
 and false-twins (in which case 
$N(u)=N(v)$).

\begin{definition}\rm \cite{Lampis} 
  A graph  $G=(V,E)$
 has \emph{neighbourhood diversity} at most $d$, if there exists a partition of $V$
 into at most $d$
 sets (we call these sets {\it type classes}) such that all the vertices in each set have the same neighbourhood type.
\end{definition}

\begin{figure}[ht]
\centering
\begin{tikzpicture}[scale=0.8]
\node[fill=dimgray,circle,draw, minimum size=0.1cm] (a) at (2, 0) [label=above:$a$]{};
\node[fill=dimgray,circle,draw, minimum size=0.1cm] (b) at (3, 0) [label=above:$b$]{};
\node[fill=dimgray,circle,draw, minimum size=0.1cm] (c) at (4, 0) [label=above:$c$]{};
\node[fill=dimgray,circle,draw, minimum size=0.1cm] (d) at (5, 0) [label=above:$d$]{};
\node[fill=red,circle,draw, minimum size=0.1cm] (e) at (3.5, -2) [label=right:$e$]{};
\node[fill=green,circle,draw, minimum size=0.1cm] (f) at (2.5, -4) [label=below :$f$]{}; 
\node[fill=green,circle,draw, minimum size=0.1cm] (g) at (1.5, -3) [label=below :$g$]{};
\node[fill=babyblue,circle,draw, minimum size=0.1cm] (h) at (5.5, -3) [label=below :$h$]{};
\node[fill=babyblue,circle,draw, minimum size=0.1cm] (i) at (4.5, -4) [label=below :$i$]{};
\node (g1) at (3.5, -5) [label=below :$G$]{}; 
\node (h1) at (11, -5) [label=below :$H$]{}; 

\path

(a) edge (e)
(b) edge (e)
(c) edge (e)
(d) edge (e)
(g) edge (e)
(f) edge (e)
(h) edge (e)
(i) edge (e)
(g) edge (f)
(g) edge (h)
(g) edge (i)
(f) edge (h)
(f) edge (i);

\node[fill=dimgray, circle,draw, minimum size=1cm ] (a0) at (11,0) []{{$a,b,c,d$}};
\node[fill=red, circle,draw, minimum size=1cm ] (b0) at (11,-2) []{{$e$}};	
\node[fill=green, circle,draw, minimum size=1cm ] (c0) at (9,-4) []{{$f,g$}};
\node[fill=babyblue, circle,draw, minimum size=1cm ] (d0) at (13,-4) []{{$h,i$}};	

\path 
(a0) edge (b0)
(b0) edge (c0)
(b0) edge (d0)
(c0) edge (d0);

\end{tikzpicture}
\caption{A graph $G$ with neighbourhood diversity 4 and its corresponding type graph $H$. }
\label{ndfig}
\end{figure}
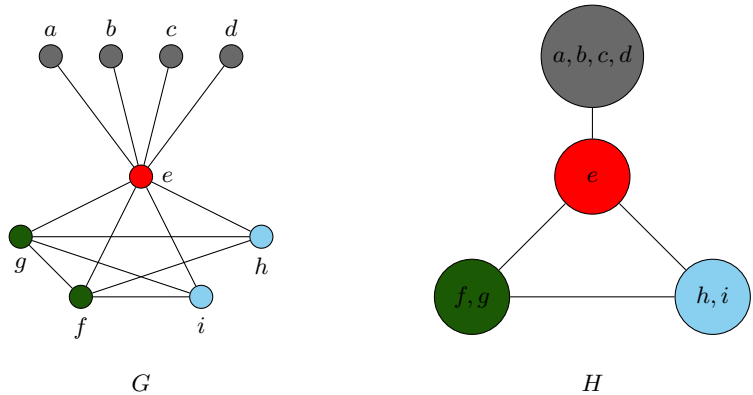

If neighbourhood diversity of a graph is bounded by an integer $d$, then there exists 
    a partition $\{ C_1, C_2,\ldots, C_d\}$ of $V(G)$ into $d$ type  classes.
    We would like to point out that it is possible to compute the neighborhood diversity of a graph in linear time
     using fast modular decomposition algorithms \cite{Tedder}. 
    Notice
    that each type class  could either be a clique or an independent set by definition and  two type classes are 
    either joined by a complete bipartite graph or no edge between vertices of the two types is present in $G$.
     For algorithmic 
    purpose it is often useful to consider a {\it type graph} $H$ of  graph $G$, where
    each vertex of $H$ is a type class in $G$,
and two vertices $C_i$ and $C_j$ are adjacent iff
    there is a complete bipartite clique between these type classes in $G$. 
     The key property of graphs of
 bounded neighbourhood diversity is that their type graphs have bounded size.  
 For example, a graph $G$ with neighbourhood diversity four and its corresponding type graph $H$ is illustrated in Figure \ref{ndfig}.
 
\noindent The following result explains why the vertices with low thresholds are inside the solution. 
\begin{lemma}\label{hslemma0}\rm 
Let $C_{i} = \{v_{1},\ldots,v_{|C_{i}|}\}$ be a type class in $G$ 
such that $t(v_{1}) \leq t(v_{2}) \leq \ldots \leq t(v_{|C_{i}|})$. 
Let $S$ be a maximum size harmless set in $G$ and $x_i=|S_i|=|C_i\cap S|$.
  Then $S' = (S\setminus S_i)\cup \{v_{1},v_{2},\ldots,v_{x_{i}}\}$ is also a maximum size harmless set in $G$.
\end{lemma}

\proof Clearly, $|S| = |S'|$.  To show $S'$ is a harmless set, it is enough
to show   that each vertex $v$ in $C_i$ has less than $t(v)$ neighbours in $S'$.  
Let $v$ be an arbitrary element of $C_i$.
 If $v\in \{v_1,\ldots, v_{x_i}\}$, 
 we have  
 \[ d_{S'}(v) = 
            \begin{cases}
               d_{S}(v) &\quad\text{if $v \in S$}\\
               d_{S}(v)-1 &\quad\text{if $v \not\in S$}\\
            \end{cases} 
            \]
Therefore, $v$ satisfies the
threshold condition $d_{S'}(v) \leq   d_{S}(v) < t(v)$.  
Suppose  $ v\in \{v_{x_{i}+1},v_{x_{i}+2}\ldots,v_{|C_{i}|}\}$. If $v\not\in S$ then $d_{S'}(v)=d_S(v) < t(v)$.  If $v\in S$ then, 
by definition of $S'$,   
some vertex $v'\in \{v_{1},v_{2},\ldots,v_{x_{i}}\}\backslash S$ must have 
replaced  $v$  as $t(v')\leq t(v)$. We have $d_{S}(v')= d_{S}(v)+1$ and also $d_{S'}(v)= d_{S}(v)+1$. It implies that $d_{S'}(v) = d_{S}(v)+1 = d_{S}(v') < t(v') \leq t(v)$.  Therefore, $S'$ is a harmless set.\\

 \noindent In this section, we prove the following theorem:
\begin{theorem}\label{theoremnd1}\rm
        The {\sc Harmless Set} problem with general thresholds  is FPT when parameterized by the neighbourhood diversity.
    \end{theorem}
    
Given a graph $G=(V,E)$ with neighbourhood diversity $nd(G)\leq d$, we first find a partition of the vertices into at most $d$ type classes $ C_1,\ldots, C_d$.  Let 
$\mathcal{C}$ be the set of all clique type classes and $\mathcal{I}$ be the set 
of all independent type classes. The case where some $C_i$ are singletons can be considered as cliques or independent sets. For simplicity, we consider singleton type classes as 
independent sets. \\
      
\noindent  {\bf ILP formulation:} Our goal here is 
to find a largest harmless set  $S$ of  $G$. 
For each $C_i$, we associate a variable $x_i$ that indicates
$|S\cap C_i|=x_i$. As the vertices in $C_i$ have the same neighbourhood, the variables $x_i$ determine $S$ uniquely, up to isomorphism. 
The threshold $t(C_i)$ of a type class $C_i$ is defined to be 
$$t(C_i)=\min \{t(v)~|~v\in C_i\}.$$  Let $\alpha(C_i)$ be the number of vertices in $C_i$ 
with threshold value $t(C_i)$.      
We define $\mathcal{C}_1=\{ C_i \in \mathcal{C}~|~ x_i<\alpha(C_i)\}$
and $\mathcal{C}_2=\{ C_i \in \mathcal{C}~|~x_i\geq \alpha(C_i)\}$. 
We next guess if a clique type class $C_i$ belongs to $\mathcal{C}_1$ or $\mathcal{C}_2$.
There are at most $2^d$ guesses as each clique type class $C_i$ has two options: either it is in 
    $\mathcal{C}_1$ or in $\mathcal{C}_2$. We reduce the problem of finding a maximum 
    harmless set
    to at most $2^d$  integer linear programming  problems with $d$ variables. 
    Since integer linear programming is fixed-parameter tractable when parameterized by 
    the number of variables \cite{lenstra}, we conclude that our problem is FPT when parameterized by 
    the neighbourhood diversity $d$. We consider the following cases based on whether $C_{i}$ is in $\mathcal{I}, \mathcal{C}_{1}$ or $\mathcal{C}_{2}$:\\

\noindent{\it Case 1:} Assume $C_i$ is in $\mathcal{I}$.

\begin{lemma}\label{hslemma1}\rm 
Let $C_i$ be an independent type class and $x_i\in \{0,1,\ldots, |C_i|\}$. Let $u_0$ be a vertex in $C_i$ with threshold
$t(C_i)$.  Then every vertex $u$ in $C_i$ has less than $t(u)$ neighbours in $S$ if and only if $u_0$ has less than $t(C_i)$ neighbours in $S$. 
\end{lemma}
\proof  Suppose each $u\in C_i$ has less than $t(u)$ neighbours in $S$.
 Then obviously $u_0 \in C_i$ has less than $t(u_0)=t(C_i)$ neighbours in $S$.  
 Conversely, suppose $u_0$ has less than $t(C_i)$ neighbours in $S$.
Let $u$ be an arbitrary vertex of $C_i$. As $u$ and $u_0$ are two vertices 
in the same type class $C_i$, we have $d_S(u)=d_S(u_0)$. 
Moreover, for each $u\in C_i$, we have  $t(C_i)\leq t(u)$ by definition of  $t(C_i)$. Therefore, 
$d_S(u)=d_S(u_0)<t(C_i)\leq t(u)$. \qed\\

\noindent Here $d_S(u_0)= \sum\limits_{C_j\in N_H(C_i)}{x_j}$. By Lemma \ref{hslemma1}, every vertex 
$u$ in $C_i$ has less than $t(u)$ neighbours in $S$ if and only if 
$$\sum\limits_{C_j\in N_H(C_i)}{x_j} < t(C_i). $$

\noindent{\it Case 2:} Assume $C_i$ is in $\mathcal{C}_1$. That is, $C_i$ is a clique
type class and $x_i< \alpha(C_i)$. Assuming $x_i< \alpha(C_i)$ ensures that 
there exists at least one vertex in $S^c\cap C_i$ with threshold $t(C_i)$.

\begin{lemma}\label{hslemma2}\rm 
Let $C_i \in \mathcal{C}_1$ and $u_0$ 
be a vertex in $S^c\cap C_i$ with threshold $t(C_i)$. Then every vertex $u$ in $C_i$ has less than $t(u)$ neighbours in $S$ if and only if  $u_0$ has less than $t(C_i)$ neighbours in $S$.  
\end{lemma}
\proof  Suppose every vertex $u$ in $C_i$ has less than $t(u)$ neighbours in $S$.
 Then obviously $u_0$ has less than $t(u_0)=t(C_i)$ neighbours in $S$.  
 Conversely, suppose $u_0$ has less than $t(C_i)$ neighbours in $S$.
Let $u$ be an arbitrary vertex of $C_i$. If  $u\in S \cap C_{i}$, then 
Lemma \ref{hslemma0} and the condition $x_i< \alpha(C_i)$ ensure  
$u$ has threshold $t(C_i)$.
Note that  $d_S(u) = d_S(u_0)-1 < t(C_i)-1 < t(C_i)=t(u)$. 
If  $ u\in S^c\cap C_{i}$, then we have $d_S(u) = d_S(u_0) < t(C_i) \leq t(u)$.
Therefore, every vertex in $C_i$ satisfies the threshold condition. \qed\\

\noindent Here $d_S(u_0)=x_i+ \sum\limits_{C_j\in N_H(C_i)}{x_j}$. By Lemma \ref{hslemma2}, every vertex 
$u$ in $C_i$ has less than $t(u)$ neighbours in $S$ if and only if 
$$x_i+\sum\limits_{C_j\in N_H(C_i)}{x_j} < t(C_i). $$

\noindent{\it Case 3:} Assume that $C_i$ is in $\mathcal{C}_2$.
 That is, $C_i$ is a clique
type class and $x_i\geq  \alpha(C_i)$. By Lemma \ref{hslemma0}, 
 all the vertices with threshold  $t(C_i)$  are inside $S$.

\begin{lemma}\label{hslemma3}\rm Let $C_i \in \mathcal{C}_2$ and $u_0$ 
be a vertex in $S\cap C_i$ with threshold $t(C_i)$.
Then every vertex $u$ in $C_i$ has less than $t(u)$ neighbours in $S$ if and only if
$u_0$ has less than $t(C_i)$ neighbours in $S$.
\end{lemma}
\proof Suppose every vertex $u$ in $C_i$ has less than $t(u)$ neighbours in $S$.
 Then obviously $u_0$ has less than $t(u_0)=t(C_i)$ neighbours in $S$.  
 Conversely,  suppose $u_0$ has less than $t(C_i)$ neighbours in $S$.
Let $u$ be an arbitrary vertex of $C_i$. If $u\in S \cap C_{i}$, then $d_S(u)=d_S(u_0)<t(C_i)\leq t(u)$. Suppose $u\in S^c\cap C_{i}$. Note that such an 
element $u$ may not always
exist, it is possible that all vertices in $C_i$ are included in $S$ (that is, $x_i=|C_i|$). Let us assume that such $u$  exists. 
Since $u$ is outside the solution and by Lemma \ref{hslemma0}, all the vertices with threshold $t(C_{i})$ are inside the solution, we get $t(u) \geq t(C_i)+1$.  It is easy to note that
 $d_S(u) = d_S(u_0)+1 < t(C_i)+1 \leq t(u)$.  \qed\\
 
 \noindent Here $d_S(u_0)=(x_i-1)+ \sum\limits_{C_j\in N_H(C_i)}{x_j}$. By Lemma \ref{hslemma3}, every vertex 
$u$ in $C_i$ has less than $t(u)$ neighbours in $S$ if and only if 
$$(x_i-1)+\sum\limits_{C_j\in N_H(C_i)}{x_j} < t(C_i). $$

\noindent The next lemma follows readily from the three lemmas above and the definition of 
 the sequence $(x_1,x_2,\ldots,x_d)$ and the harmless set. 

\begin{lemma}\rm 
Let $G=(V,E)$ be a graph such that $V$ can be partitioned into at most $d$ type 
classes $ C_1,\ldots, C_d$. The sequence $(x_1,x_2,\ldots,x_d)$ represents a 
harmless set $S$ of $G$ if and only if $(x_1,x_2,\ldots,x_d)$ satisfies 
\begin{enumerate}
 \item  $x_i\in \{0,1,\ldots, |C_i|\} ~~\text{for } i=1,2,\ldots,d$
    \item $\sum\limits_{C_j\in N_H(C_i)}{x_j} < t(C_i)$ for all  $C_i \in \mathcal{I}$.
    \item $x_i+\sum\limits_{C_j\in N_H(C_i)}{x_j} < t(C_i)$ and $x_i< \alpha(C_i)$ for all $C_i \in \mathcal{C}_1$ 
    \item $(x_i-1)+\sum\limits_{C_j\in N_H(C_i)}{x_j} < t(C_i)$ and $ \alpha(C_i) \leq x_i\leq |C_i|$ for all $C_i\in \mathcal{C}_2$.
\end{enumerate}

\end{lemma}

\noindent In the following, we present an ILP formulation for the
        {\sc Harmless Set} problem parameterized by neighbourhood 
        diversity for a guess:\\
        
\noindent       
\vspace{3mm}
    \fbox
    {\begin{minipage}{33.7em}\label{Min-FFVS0}        
\begin{equation*}
\begin{split}
&\text{Maximize} \sum\limits_{C_i}{x_i} \\
&\text{Subject to~~~} \\
& x_i\in \{0,1,\ldots, |C_i|\} ~~\text{for } i=1,2,\ldots,d\\
& \sum\limits_{C_j\in N_H(C_i)}{x_j} < t(C_i), ~~\text{for all } C_i\in \mathcal{I}, \\
& x_i+\sum\limits_{C_j\in N_H(C_i)}{x_j} < t(C_i)~~ \mbox{and } x_i< \alpha(C_i) \text{ for all } 
C_i\in \mathcal{C}_1 \\
&  (x_i-1)+\sum\limits_{C_j\in N_H(C_i)}{x_j} < t(C_i) \mbox{ and } 
\alpha(C_i) \leq x_i\leq |C_i| ~~\text{for all } C_i\in \mathcal{C}_2
\end{split}
\end{equation*}  
  \end{minipage} }
    \vspace{3mm}
    
\begin{example}\label{ndexample}\rm 
Consider a graph composed of a clique  $C$ of size  $c+1 ~(\geq 4)$ plus a vertex $u$
adjacent to a vertex $v$ of the clique as shown in Figure \ref{ndexample1}. We set unanimity thresholds, so $t(u)=1$, $t(v)=c+1$ and $t(x)=c$ for all $x$ in $C\setminus \{v\}$.
\begin{figure}[ht]
\centering
\begin{tikzpicture}[scale=0.7]
\node[circle,draw, minimum size=.1cm ] (x) at (0,2) []{{$4$}};
\node[circle,draw, minimum size=.1cm ] (y) at (0,0) []{{$4$}};
\node[circle,draw, minimum size=.1cm ] (w) at (2,2) []{{$4$}};
\node[circle,draw, minimum size=.1cm ] (z) at (2,0) []{{$4$}};
\node[circle,draw, minimum size=.1cm ] (v) at (3,1) [label=above:$v$]{{$5$}};
\node[circle,draw, minimum size=.1cm ] (u) at (5,1) [label=above:$u$]{{$1$}};

\path 
(u) edge (v)
(v) edge (w)
(v) edge (x)
(v) edge (y)
(v) edge (z)
(w) edge (x)
(w) edge (y)
(w) edge (z)
(x) edge (y)
(x) edge (z)
(y) edge (z)
;

\end{tikzpicture}
\caption{The graph in Example  \ref{ndexample} with $c=4$. }
\label{ndexample1}
\end{figure}
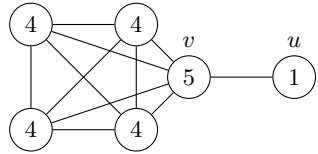
The type classes are $C_1=\{u\}$, $C_2=\{v\}$ and $C_3=C\setminus \{v\}$.
Here $\alpha(C_1)=1$, $\alpha(C_2)=1$ and $\alpha(C_3)=|C_3|=c.$
Now consider the guess  $C_1,C_2\in \mathcal {I}$ and 
$C_3\in \mathcal{C}_2$. Then we end up with the following ILP:
\begin{align*}
    \text{min} \quad
    & x_1+x_2+x_3  \\
    \text{s.t.} \quad
    &x_2 <1\\
    &x_1+x_3 <c+1\\
    &x_3-1+x_2<c \mbox{ and } x_3=c
\end{align*}
Note that $x_2<1$ implies  that $x_2=0$; $x_1+x_3 <c+1$ and $x_3=c$ imply $x_1=0$. 
It is easy to see that $x_1=x_2=0$, $x_3=c$ is an optimal solution and represent a valid harmless set for the graph. 

\end{example}    
    
\noindent {\bf Solving the ILP}
Lenstra \cite{lenstra} showed that the feasibility version of {\sc $p$-ILP} is FPT with 
running time doubly exponential in $p$, where $p$ is the number of variables. 
Later, Kannan \cite{kannan} proved an algorithm for {\sc $p$-ILP} running in time $p^{O(p)}$.
In our algorithm, we need the optimization version of {\sc $p$-ILP} rather than 
the feasibility version. We state the minimization version of {\sc $p$-ILP}
as presented by Fellows et. al. \cite{fellows}. \\

\noindent {\sc $p$-Variable Integer Linear Programming Optimization ($p$-Opt-ILP)}: Let matrices $A\in \ Z^{m\times p}$, $b\in \ Z^{p\times 1}$ and 
$c\in \ Z^{1\times p}$ be given. We want to find a vector $ x\in \ Z ^{p\times 1}$ that minimizes the objective function $c\cdot x$ and satisfies the $m$ 
inequalities, that is, $A\cdot x\geq b$.  The number of variables $p$ is the parameter. 
Then they showed the following:

\begin{proposition}\rm \label{ilp}\cite{fellows}
{\sc $p$-Opt-ILP} can be solved using $O(p^{2.5p+o(p)}\cdot L \cdot \log(MN))$ arithmetic operations and space polynomial in $L$. 
Here $L$ is the number of bits in the input, $N$ is the maximum absolute value any variable can take, and $M$ is an upper bound on 
the absolute value of the minimum taken by the objective function.
\end{proposition}

In the formulation for {\sc Harmless Set} problem, we have at most $d$ variables. The value of the objective function is bounded by $n$ and the value of any variable 
in the integer linear programming is also bounded by $n$. The constraints can be represented using 
$O(d^2 \log{n})$ bits. Proposition \ref{ilp} implies that we can solve the problem with the guess $\mathcal{P}$ in FPT time. 
There are at most $2^d$ guesses, and the ILP formula for a guess can be solved in FPT time. Thus 
Theorem \ref{theoremnd1} holds. 

\section{FPT algorithm parameterized by twin cover}
In this section, we present an FPT algorithm for the {\sc Harmless Set} problem with 
general thresholds parameterized by twin-cover. 
That is, we prove the following theorem:
\begin{theorem}\label{theorem-tc}\rm
 The {\sc Harmless Set} problem with general thresholds is FPT when parameterized by twin cover of the input graph.
 \end{theorem}
 
\noindent{ \it Outline of the algorithm.} Given an $n$-vertex graph $G$ with ${\tt tc}(G) \leq  k$, we first find a twin cover $X$ of size at most $k$. 
We next guess $S_X=S\cap X$ where $S$ is a largest harmless set in $G$.
There are at most $2^k$ guesses as each member of $X$ has two options:
either in $S\cap X$ or $S^c \cap X$.  Finally we reduce the problem of finding the 
rest of $S$ to an integer linear programming (ILP) optimization with 
at most $2^k$ variables. Since ILP optimization is fixed-parameter tractable when parameterized by the number of variables \cite{fellows}, we can conclude that our problem is fixed-parameter tractable when parameterized by the twin cover number.\\

\noindent{\it Characterizations of a harmless set S with a twin cover X.}
Let $G = (V,E)$ be a graph and $X\subseteq  V$ be a twin cover of $G$.
Then $\mathcal{C}=G\setminus X$
is a collection of disjoint cliques, that is  $\mathcal{C}=\{C_1,C_2,\ldots\}$. 
The threshold $t(C_i)$ of a clique $C_i$ is defined to be 
$$t(C_i)=\min \{t(v)~|~v\in C_i\}.$$  Let $\alpha(C_i)$ be the number of vertices in $C_i$ 
with threshold value $t(C_i)$. It may be observed that from a clique 
it is always better to include the vertices with lower thresholds in the solution.  
The reason is this. 
Suppose $a$ and $b$ are two vertices from the same clique $C$ such that
$t(a)<t(b)$.
Suppose $a$ is in the solution, whereas $b$ is outside the solution. 
Then $a$ has one less neighbours in the solution compare to $b$. 
 This one less degree of $a$ in the solution  helps the vertex $a$ 
to satisfy the required threshold condition as $t(a)<t(b)$.      
We define $\mathcal{C}_{>}=\{ C \in \mathcal{C}~:~\alpha(C)>t(C)-|N_{S_{X}}(C)| \}$
and $\mathcal{C}_{\leq}=\{C\in \mathcal{C}~:~\alpha(C)\leq t(C)-|N_{S_{X}}(C)|\}$. 
 \\


\begin{lemma}\label{hslemma5}\rm 
Assume that  $C$ is in $\mathcal{C}_>$. Then every vertex $u$ in $C$ has less than $t(u)$ neighbours in $S$ if and only if $|S \cap C|\leq t(C) - |N_{S_{X}}(C)|-1$.  
\end{lemma}

\proof Suppose every vertex $u$ in $C$ has less than $t(u)$ neighbours in $S$, and suppose, for the sake of contradiction, that $|S \cap C|\geq  t(C) - |N_{S_{X}}(C)|$. 
If  $|S \cap C|= t(C) - |N_{S_{X}}(C)|$ then $ |S\cap C|< \alpha(C)$. This implies there exists a vertex $u_{0}\in S^{c}\cap C$ with $t(u_{0})=t(C)$. 
Note that $d_S(u_0)= |S\cap C|+ |N_{S_X}(C)|=t(C)=t(u_0)$, a contradiction
to the assumption that every vertex in $C$ has less than $t(u)$ neighbours in $S$. 
Let us assume that $|S \cap C|\geq t(C) - |N_{S_{X}}(C)|+1$. Let $u$ be an 
arbitrary vertex in $C$. Then $d_S(u)\geq |S\cap C|-1+|N_{S_X}(C)| \geq t(C)\geq t(u)$,
which is again a contradiction. This proves the forward direction. \\
On the other hand,  let us assume that $|S \cap C|\leq t(C) - |N_{S_{X}}(C)|-1$. It implies that $ |S\cap C| < \alpha(C)$ and this ensures that 
there exists at least one vertex $u_0$ in $S^c\cap C$ with threshold $t(C)$.
By Lemma \ref{hslemma2}, it is enough to check whether $u_{0}\in S^{c}\cap C$ with threshold $t(C)$ satisfies the threshold condition. Clearly, $d_{S}(u_{0}) = |S\cap C|+ |N_{S_X}(C)| \leq t(C)-|N_{S_{X}}(C)|-1+|N_{S_X}(C)|= t(C)-1$ satisfies the threshold condition. Therefore, every vertex $u$ in $C$ has less than $t(u)$ neighbours in $S$. \qed\\

\begin{lemma}\label{hslemma6}\rm 
Assume that   $C$ is in  $ \mathcal{C}_{\leq}$. Then every vertex $u$ in $C$ has less than $t(u)$ neighbours in $S$ if and only if $|S \cap C|\leq t(C) - |N_{S_{X}}(C)|$.  
\end{lemma}

\proof  
 Suppose each vertex $u\in C$ has less than $t(u)$ neighbours in $S$.
 Let $u_0$ be a vertex in $C$ with threshold $t(u_0)=t(C)$. Then $u_0$ also 
 has less than $t(u_0)$ neighbours in $S$, that is, 
$d_S(u_0) \leq |S\cap C|+N_{S_X}(C) < t(u_0)=t(C)$. Therefore, we get
$|S \cap C|\leq t(C) - |N_{S_{X}}(C)|$.\\
On the other hand, first suppose that $|S \cap C|= t(C) - |N_{S_{X}}(C)|$. Therefore, we can say that $ |S\cap C|\geq \alpha(C)$. It means that all the vertices with least threshold are inside the solution. 
 Let $u$ be an arbitrary vertex in $S \cap C$ with threshold $t(u)$. By Lemma \ref{hslemma3}, it is enough to check whether $u$ satisfies the threshold condition.
 It is easy to observe that
 $d_{S}(u)=|S\cap C|-1+|N_{S_X}(C)|= t(C)-1<t(C)=t(u)$, that is, $u$ satisfies
 the threshold condition. Now suppose $|S \cap C|= t(C) - |N_{S_{X}}(C)|-\delta$ 
 for some integer $\delta\geq 1$. Take an arbitrary vertex $u\in C$. We have $d_{S}(u)\leq |S \cap C|+ |N_{S_{X}}(C)| = t(C)-\delta <t(C)\leq t(u)$. This implies that all the vertices in $C$ satisfy the threshold condition.  \qed\\

\noindent We partition the family $\mathcal{C}$ of cliques into  twin classes 
$\mathcal{C}_1, \mathcal{C}_2, \ldots, 
\mathcal{C}_t$, where $t\leq 2^k$. Two cliques $C_i$ and $C_j$ are in the same 
twin class if and only if they have the same neighbours in $X$, that is, $N_X(C_i)=N_X(C_j)$.
For each twin class $\mathcal{C}_i$, we associate a variable $x_i$ that indicates 
$|\mathcal{C}_i\cap S|=x_i$.  The variables $x_i$ determine $S$. The objective is to maximize $\sum\limits_{i=1}^{t}{x_i}$
under the condition 
$$x_{i}\leq \sum\limits_{C\in \mathcal{C}_> \cap \mathcal{C}_i}\Big(t(C)-|N_{S_X}(C)|-1 \Big) + \sum\limits_{C\in \mathcal{C}_\leq \cap \mathcal{C}_i} \Big( t(C)-|N_{S_X}(C)|\Big).$$
The above constraint makes sure that the vertices of twin class $\mathcal{C}_i$  satisfy 
the threshold condition. Next, we add $k$ more constraints to make sure that all the vertices in $X$ satisfy threshold condition. 
Let $u\in X$ be an arbitrary vertex. We need $d_S(u) < t(u)$. Note that
$d(u)  = d_X(u)+ \sum\limits_{\mathcal{C}_i ~:~ N(u)\cap \mathcal{C}_i\neq \emptyset} |\mathcal{C}_i|$. Therefore, for each $u\in X$, we add the constraint 
$d_S(u)=d_{S_X}(u)+ \sum\limits_{\mathcal{C}_i ~:~ N(u)\cap \mathcal{C}_i\neq \emptyset} {x_i} < t(u)$. 

        \noindent In the following, we present an ILP formulation for the
        {\sc Harmless Set} problem for a given $S_X$:\\
        
\noindent       
\vspace{3mm}
    \fbox
    {\begin{minipage}{35.7em}\label{Min-FFVS}        
\begin{equation*}
\begin{split}
&\text{Maximize} \sum\limits_{i=1}^{t}{x_i}, ~t\leq 2^k \\
&\text{Subject to~~~} \\
&x_{i}\leq \sum\limits_{C\in \mathcal{C}_> \cap \mathcal{C}_i}\Big(t(C)-|N_{S_X}(C)|-1 \Big) + \sum\limits_{C\in \mathcal{C}_\leq \cap \mathcal{C}_i} \Big( t(C)-|N_{S_X}(C)|\Big), ~\text{for~} 1\leq i \leq 2^{k} \\
&  d_{S_X}(u) +\sum\limits_{\mathcal{C}_i ~:~ N(u)\cap \mathcal{C}_i\neq \emptyset} {x_i} < t(u)~~\text{for all } 
u\in X 
\end{split}
\end{equation*}  
  \end{minipage} }
    \vspace{3mm}
    
\noindent In the ILP formulation for {\sc Harmless Set} problem parameterized by twin cover, we have at most $2^k$ variables. The value of objective function is bounded by $n$ and the value of any variable 
in the integer linear programming is also bounded by $n$. The constraints can be represented using 
$O(k^2 \log{n})$ bits. Proposition \ref{ilp} implies that we can solve the problem
for given $S_X$ in FPT time. 
There are at most $2^k$ guesses for $S_X$, and the ILP formula for a guess can be solved in FPT time. Thus 
Theorem \ref{theorem-tc} holds. 

\section{FPT algorithm parameterized by vertex integrity}

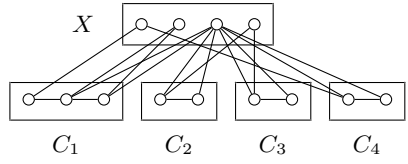
\begin{figure}[ht]
    \centering
    \begin{tikzpicture}[scale=0.5]
\node[circle, draw, inner sep=0 pt, minimum size=0.15cm](x1) at (0,0) [label=above:]{};
\node[circle, draw, inner sep=0 pt, minimum size=0.15cm](x2) at (1,0) [label=above:]{};
\node[circle, draw, inner sep=0 pt, minimum size=0.15cm](x3) at (2,0) [label=above:]{};
\node[circle, draw, inner sep=0 pt, minimum size=0.15cm](x5) at (3,0) [label=above:]{};

\node(x) at (-.8,0) [label=left:$X$]{};

\node[circle, draw, inner sep=0 pt, minimum size=0.15cm](a1) at (-3,-2) [label=above:]{};
\node[circle, draw, inner sep=0 pt, minimum size=0.15cm](a2) at (-2,-2) [label=above:]{};
\node[circle, draw, inner sep=0 pt, minimum size=0.15cm](a3) at (-1,-2) [label=above:]{};

\node(k1) at (-2,-2.5) [label=below:$C_{1}$]{};
\node[circle, draw, inner sep=0 pt, minimum size=0.15cm](b1) at (0.5,-2) [label=above:]{};
\node[circle, draw, inner sep=0 pt, minimum size=0.15cm](b2) at (1.5,-2) [label=above:]{};

\node(k2) at (1,-2.5) [label=below:$C_{2}$]{};
\node[circle, draw, inner sep=0 pt, minimum size=0.15cm](c1) at (3,-2) [label=above:]{};
\node[circle, draw, inner sep=0 pt, minimum size=0.15cm](c2) at (4,-2) [label=above:]{};

\node(k3) at (3.5,-2.5) [label=below:$C_{3}$]{};
\node[circle, draw, inner sep=0 pt, minimum size=0.15cm](d1) at (5.5,-2) [label=above:]{};
\node[circle, draw, inner sep=0 pt, minimum size=0.15cm](d2) at (6.5,-2) [label=above:]{};

\node(k4) at (6,-2.5) [label=below:$C_{4}$]{};

\node[circle, draw, inner sep=0 pt, minimum size=0.005cm](s11) at (-.5,.55) [label=below:]{};
\node[circle, draw, inner sep=0 pt, minimum size=0.005cm](s12) at (-.5,-.55) [label=below:]{};
\node[circle, draw, inner sep=0 pt, minimum size=0.005cm](s13) at (3.5,.55) [label=below:]{};
\node[circle, draw, inner sep=0 pt, minimum size=0.005cm](s14) at (3.5,-.55) [label=below:]{};

\node[circle, draw, inner sep=0 pt, minimum size=0.005cm](s21) at (-3.5,-1.55) [label=below:]{};
\node[circle, draw, inner sep=0 pt, minimum size=0.005cm](s22) at (-3.5,-2.55) [label=below:]{};
\node[circle, draw, inner sep=0 pt, minimum size=0.005cm](s23) at (-.5,-1.55) [label=below:]{};
\node[circle, draw, inner sep=0 pt, minimum size=0.005cm](s24) at (-.5,-2.55) [label=below:]{};

\node[circle, draw, inner sep=0 pt, minimum size=0.005cm](s31) at (2,-1.55) [label=below:]{};
\node[circle, draw, inner sep=0 pt, minimum size=0.005cm](s32) at (2,-2.55) [label=below:]{};
\node[circle, draw, inner sep=0 pt, minimum size=0.005cm](s33) at (0,-1.55) [label=below:]{};
\node[circle, draw, inner sep=0 pt, minimum size=0.005cm](s34) at (0,-2.55) [label=below:]{};

\node[circle, draw, inner sep=0 pt, minimum size=0.005cm](s41) at (2.5,-1.55) [label=below:]{};
\node[circle, draw, inner sep=0 pt, minimum size=0.005cm](s42) at (2.5,-2.55) [label=below:]{};
\node[circle, draw, inner sep=0 pt, minimum size=0.005cm](s43) at (4.5,-1.55) [label=below:]{};
\node[circle, draw, inner sep=0 pt, minimum size=0.005cm](s44) at (4.5,-2.55) [label=below:]{};

\node[circle, draw, inner sep=0 pt, minimum size=0.005cm](s51) at (5,-1.55) [label=below:]{};
\node[circle, draw, inner sep=0 pt, minimum size=0.005cm](s52) at (5,-2.55) [label=below:]{};
\node[circle, draw, inner sep=0 pt, minimum size=0.005cm](s53) at (7,-1.55) [label=below:]{};
\node[circle, draw, inner sep=0 pt, minimum size=0.005cm](s54) at (7,-2.55) [label=below:]{};

\draw(s51)--(s52);
\draw(s54)--(s52);
\draw(s53)--(s51);
\draw(s53)--(s54);

\draw(s41)--(s42);
\draw(s44)--(s42);
\draw(s43)--(s41);
\draw(s43)--(s44);

\draw(s31)--(s32);
\draw(s34)--(s32);
\draw(s33)--(s31);
\draw(s33)--(s34);

\draw(s21)--(s22);
\draw(s24)--(s22);
\draw(s23)--(s21);
\draw(s23)--(s24);

\draw(s11)--(s12);
\draw(s14)--(s12);
\draw(s13)--(s11);
\draw(s13)--(s14);

\draw(x1)--(a1);
\draw(x3)--(a3);
\draw(x3)--(a2);
\draw(x3)--(b1);
\draw(x3)--(b2);
\draw(x3)--(c1);
\draw(x3)--(c2);
\draw(x3)--(d1);
\draw(x3)--(d2);
\draw(x5)--(b1);
\draw(x5)--(c1);

\draw(a1)--(a2);
\draw(a2)--(a3);
\draw(b2)--(b1);
\draw(c2)--(c1);
\draw(d2)--(d1);
\draw(x2)--(a2);
\draw(x2)--(a3);
\draw(x1)--(d1);

    \end{tikzpicture}
    \caption{The components $C_{2}$ and $C_{3}$ of $G-X$ have the same $(G,X)-$type.}
    \label{videf}
\end{figure}

In this section, we present an FPT algorithm for {\sc Harmless Set} parameterized by vertex integrity.
\begin{definition}\rm 
The \emph{vertex integrity} of a graph $G$, denoted ${\tt vi}(G)$, is the minimum integer $k$ satisfying that there is  $X \subseteq V (G)$ such that  $|X|+|V(C)| \leq  k$ for each component 
$C$ of  $G-X$. We call such $S$ a ${\tt vi}(k)$-set of $G$.  
\end{definition}
\noindent This parameter is bounded from above by vertex cover number plus one and from below by treedepth.  Gima et al. \cite{Gima} described an 
equivalence relation among components. 
For a vertex set $X$ of $G$, we define an equivalence relation $\sim_{G,X}$ among components of $G-X$ by setting $C_1 \sim_{G,X} C_2$ if and only 
if there is an isomorphism $g$ from $G[X \cup  V(C_1)] $ to $G[X \cup V(C_2)]$ 
that fixes $X$; that is, $g|_X$ is the identity function. 
When $ C_1 \sim_{G,X} C_2$, we say that $C_1$ and $C_2$ have the same $(G,X)$-\emph{type} (or just the same \emph{type} if $G$ and $X$ are clear from the context). 
See Figure \ref{videf}.  
This equivalence relation induces a set of equivalence classes $\mathcal{C}_1,
\mathcal{C}_2,\ldots$. We can choose a representative of each equivalence class.



\begin{theorem}\label{theorem-vi} The {\sc Harmless Set} problem is fixed-parameter tractable when parameterized by the vertex integrity of the input graph.
\end{theorem}

\noindent{\it Outline of the algorithm.} Let $X$ be a ${\tt vi}(k)$-set of $ G$. Such a set can be found in $O(k^{k+1}n)$ time \cite{Pal}. We next guess $S_X=S\cap X$ where $S$ is a largest harmless set in $G$.
There are at most $2^k$ guesses as each member of $X$ has two options:
either in $S\cap X$ or $S^c \cap X$. Finally we reduce the problem of finding the 
rest of $S$ to an integer linear programming (ILP) optimization with 
number of variables depend only on $k$. \\

\noindent{\it Characterizations of a harmless set $S$ with a {\tt vi}$(k)$-set  $X$.}
Let $G = (V,E)$ be a graph and $X\subseteq  V$ be a {\tt vi}$(k)$-set of $G$.
Then $\mathcal{C}=G\setminus X$
is a collection of disjoint components, that is  $\mathcal{C}=\{C_1,C_2,\ldots\}$
such that $|X|+|C_i|\leq k$ for all $i$.  We know $\mathcal{C}$ can be partitioned into 
equivalent classes $\mathcal{C}_1, \mathcal{C}_2,\ldots$. Let $C_{l}$ be a 
representative of the equivalence 
class $\mathcal{C}_l$ and let $v\in C_l$.
 Note that $v$ has neighbours only in $X\cup V(C_l)$, that is, $N(v)\subseteq 
X\cup V(C_l)$. Suppose the intersection of the solution $S$ with $X$ is $S_X=S\cap X$ and the intersection of $S$ with the component $C_l$
is $A=S\cap C_l \subseteq C_l$.  Therefore  $v \in C_l$ satisfies the threshold condition 
if $d_S(v)=|N_{S_X}(v)|+|N_A(v)|< t(v)$. 
We say  $A\subseteq C_l$ is a 
\emph{valid selection} for $C_l$
 if  every vertex of $C_l$ satisfies
the threshold condition when the vertices of
$A \cup S_X$ are in the solution. Similarly, we say $A\subseteq C_l$ is a 
\emph{valid selection} for $C \in \mathcal{C}_l$, $C\neq C_l$,
 if  every vertex of $C$ satisfies
the threshold condition when the vertices of
 $g(A) \cup S_X$ are in the solution, where $g$ is an isomorphism  from $G[X \cup  V(C_l)] $ to $G[X \cup V(C)]$ 
that fixes $X$. It is important to note that 
given two connected component $C_{1}$ and $C_{2}$ from the same equivalence class
$\mathcal{C}_l$,  a subset $A\subseteq C_l$
 might be valid selection for one connected component but 
may not be valid for the other connected component as the threshold values
of vertices in $C_{1}$ and $C_{2}$ can differ.

\begin{figure}[ht]
    \centering
    \begin{tikzpicture}

\node[rectangle,
    fill = green,
    minimum width = 3cm, 
    minimum height = 1.1cm, opacity=0.5] (s) at (3,0) [label=above:$S_{X}$]{};

\node[rectangle,
    fill = purple,
    minimum width = 2cm, 
    minimum height = 1cm, opacity=0.5] (p) at (-1.5,-2.05){};
    
\node (p1) at (-1.5,-2.05)[label=below:$B$]{};
    
\node[rectangle,
    fill = orange,
    minimum width = 1cm, 
    minimum height = 1cm, opacity=0.5] (q) at (0.5,-2.05) {};

\node (q1) at (0.5,-2.05)[label=left:$A$]{};

\node[rectangle,
    fill = orange,
    minimum width = 1cm, 
    minimum height = 1cm, opacity=0.5] (r) at (3,-2.05) {};
    
\node (r1) at (3,-2.05)[label=left:$g(A)$]{};
    
 \node[rectangle,
    fill = blue,
    minimum width = 2cm, 
    minimum height = 1cm, opacity=0.5] (t) at (6,-2.05) {};
    
\node (r1) at (6,-2.05)[label=below:$C$]{};

\node[circle, draw, inner sep=0 pt, minimum size=0.15cm](x1) at (0,0) [label=above:$v_1$]{};
\node[circle, draw, inner sep=0 pt, minimum size=0.15cm](x2) at (1,0) [label=above:$v_2$]{};
\node[circle, draw, inner sep=0 pt, minimum size=0.15cm](x3) at (2,0) [label=above:$v_3$]{};
\node[circle, draw, inner sep=0 pt, minimum size=0.15cm](x4) at (3,0) [label=above:$v_4$]{};
\node[circle, draw, inner sep=0 pt, minimum size=0.15cm](x5) at (4,0) [label=above:$v_5$]{};

\node(x) at (-.8,0) [label=left:$X$]{};
\node[circle, draw, inner sep=0 pt, minimum size=0.15cm](a1) at (-3,-2) [label=above:$a$]{};
\node[circle, draw, inner sep=0 pt, minimum size=0.15cm](a1) at (-3,-2) [label=below:$1$]{};
\node[circle, draw, inner sep=0 pt, minimum size=0.15cm](a2) at (-2,-2) [label=above:$b$]{};
\node[circle, draw, inner sep=0 pt, minimum size=0.15cm](a2) at (-2,-2) [label=below:$2$]{};
\node[circle, draw, inner sep=0 pt, minimum size=0.15cm](a3) at (-1,-2) [label=above:$c$]{};
\node[circle, draw, inner sep=0 pt, minimum size=0.15cm](a3) at (-1,-2) [label=below:$3$]{};

\node(k1) at (-2,-2.5) [label=below:$C_{1}$]{};
\node[circle, draw, inner sep=0 pt, minimum size=0.15cm](b1) at (0.5,-2) [label=above:$d$]{};
\node[circle, draw, inner sep=0 pt, minimum size=0.15cm](b1) at (0.5,-2) [label=below:$3$]{};
\node[circle, draw, inner sep=0 pt, minimum size=0.15cm](b2) at (1.5,-2) [label=above:$e$]{};
\node[circle, draw, inner sep=0 pt, minimum size=0.15cm](b2) at (1.5,-2) [label=below:$3$]{};

\node(k2) at (1,-2.5) [label=below:$C_{2}$]{};
\node[circle, draw, inner sep=0 pt, minimum size=0.15cm](c1) at (3,-2) [label=above:$f$]{};
\node[circle, draw, inner sep=0 pt, minimum size=0.15cm](c1) at (3,-2) [label=below:$3$]{};
\node[circle, draw, inner sep=0 pt, minimum size=0.15cm](c2) at (4,-2) [label=above:$g$]{};
\node[circle, draw, inner sep=0 pt, minimum size=0.15cm](c2) at (4,-2) [label=below:$2$]{};

\node(k3) at (3.5,-2.5) [label=below:$C_{3}$]{};
\node[circle, draw, inner sep=0 pt, minimum size=0.15cm](d1) at (5.5,-2) [label=above:$h$]{};
\node[circle, draw, inner sep=0 pt, minimum size=0.15cm](d1) at (5.5,-2) [label=below:$4$]{};
\node[circle, draw, inner sep=0 pt, minimum size=0.15cm](d2) at (6.5,-2) [label=above:$i$]{};

\node[circle, draw, inner sep=0 pt, minimum size=0.15cm](d2) at (6.5,-2) [label=below:$3$]{};

\node(k4) at (6,-2.5) [label=below:$C_{4}$]{};

\node[circle, draw, inner sep=0 pt, minimum size=0.005cm](s11) at (-.5,.55) [label=below:]{};
\node[circle, draw, inner sep=0 pt, minimum size=0.005cm](s12) at (-.5,-.55) [label=below:]{};
\node[circle, draw, inner sep=0 pt, minimum size=0.005cm](s13) at (4.5,.55) [label=below:]{};
\node[circle, draw, inner sep=0 pt, minimum size=0.005cm](s14) at (4.5,-.55) [label=below:]{};

\node[circle, draw, inner sep=0 pt, minimum size=0.005cm](s21) at (-3.5,-1.55) [label=below:]{};
\node[circle, draw, inner sep=0 pt, minimum size=0.005cm](s22) at (-3.5,-2.55) [label=below:]{};
\node[circle, draw, inner sep=0 pt, minimum size=0.005cm](s23) at (-.5,-1.55) [label=below:]{};
\node[circle, draw, inner sep=0 pt, minimum size=0.005cm](s24) at (-.5,-2.55) [label=below:]{};

\node[circle, draw, inner sep=0 pt, minimum size=0.005cm](s31) at (2,-1.55) [label=below:]{};
\node[circle, draw, inner sep=0 pt, minimum size=0.005cm](s32) at (2,-2.55) [label=below:]{};
\node[circle, draw, inner sep=0 pt, minimum size=0.005cm](s33) at (0,-1.55) [label=below:]{};
\node[circle, draw, inner sep=0 pt, minimum size=0.005cm](s34) at (0,-2.55) [label=below:]{};

\node[circle, draw, inner sep=0 pt, minimum size=0.005cm](s41) at (2.5,-1.55) [label=below:]{};
\node[circle, draw, inner sep=0 pt, minimum size=0.005cm](s42) at (2.5,-2.55) [label=below:]{};
\node[circle, draw, inner sep=0 pt, minimum size=0.005cm](s43) at (4.5,-1.55) [label=below:]{};
\node[circle, draw, inner sep=0 pt, minimum size=0.005cm](s44) at (4.5,-2.55) [label=below:]{};

\node[circle, draw, inner sep=0 pt, minimum size=0.005cm](s51) at (5,-1.55) [label=below:]{};
\node[circle, draw, inner sep=0 pt, minimum size=0.005cm](s52) at (5,-2.55) [label=below:]{};
\node[circle, draw, inner sep=0 pt, minimum size=0.005cm](s53) at (7,-1.55) [label=below:]{};
\node[circle, draw, inner sep=0 pt, minimum size=0.005cm](s54) at (7,-2.55) [label=below:]{};

\draw(s51)--(s52);
\draw(s54)--(s52);
\draw(s53)--(s51);
\draw(s53)--(s54);

\draw(s41)--(s42);
\draw(s44)--(s42);
\draw(s43)--(s41);
\draw(s43)--(s44);

\draw(s31)--(s32);
\draw(s34)--(s32);
\draw(s33)--(s31);
\draw(s33)--(s34);

\draw(s21)--(s22);
\draw(s24)--(s22);
\draw(s23)--(s21);
\draw(s23)--(s24);

\draw(s11)--(s12);
\draw(s14)--(s12);
\draw(s13)--(s11);
\draw(s13)--(s14);

\draw(x1)--(a1);
\draw(x3)--(a3);
\draw(x3)--(a2);
\draw(x3)--(b1);
\draw(x3)--(b2);
\draw(x3)--(c1);
\draw(x3)--(c2);
\draw(x3)--(d1);
\draw(x3)--(d2);
\draw(x4)--(d1);
\draw(x5)--(b1);
\draw(x5)--(c1);

\draw(a1)--(a2);
\draw(a2)--(a3);
\draw(b2)--(b1);
\draw(c2)--(c1);
\draw(d2)--(d1);

    \end{tikzpicture}
    \caption{The components $C_{2}$ and $C_{3}$ of $G-X$ have the same $(G,X)-$type. That is, $C_{2}$ and $C_{3}$ are in the same equivalence class. 
    Note that given $S_{X}=\{v_3,v_4,v_5\}$, $A=\{d\}$ is a valid selection for $C_{2}$ but $g(A)=\{f\}$ is not a valid selection for $C_{3}$; $B$ is not a valid selection for $C_{1}$; $C$ is a valid selection for $C_{4}$.}
    \label{vi-example}
\end{figure}
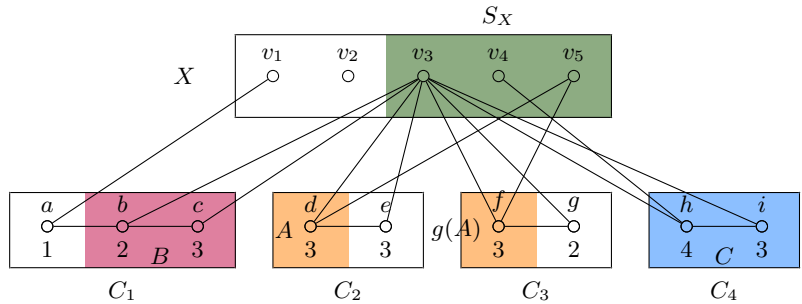

Given $S_X=S\cap X$, for every equivalence class $\mathcal{C}_l$, we define
the set of valid selection
$$ \mathcal{VS}(l)=\Big\{ A\subseteq  C_l~|~ A \mbox{ is a valid selection for 
some } C\in \mathcal{C}_l\Big \}. $$
We denote by $\rho(A)$ the number of components 
in $\mathcal{C}_l$  where
$A$ is a valid selection.
Similarly, we denote by $\rho(A_{i_1}A_{i_2}\ldots A_{i_r})$  the number of components 
in $\mathcal{C}_l$  where $A_{i_1},A_{i_2},\ldots, A_{i_r}$ are valid 
selection.  Note that for each component there may be more than 
one valid selections.  But while forming the harmless set we can pick at
most one valid selection for every component.
Observe that as long as we are taking exactly one valid selection for each connected component, we are guaranteed that all the vertices in $G-X$ satisfy the threshold conditions.  We  need to add constraints in ILP separately to make sure that every vertex
of $X$ also satisfies the threshold condition.  Let $x(A)$ denote the number of components for which 
$A$ has been picked as a valid selection in the final solution. They satisfy the following properties:
\begin{equation*}
\begin{split}
    x(A_i) & \leq \rho(A_i) ~\mbox{ for all }~ A_i\in \mathcal{VS}(l)\\
    x(A_i)+x(A_j) &\leq \rho(A_i)+\rho(A_j)-\rho(A_iA_j)~\mbox{ for all }~ A_i,A_j\in \mathcal{VS}(l)\\
    &\vdots\\
    \sum_{A_i\in \mathcal{VS}(l)}{x(A_i)}&= |\mathcal{C}_l|
\end{split}
\end{equation*}

\begin{example}
Suppose the equivalent class $\mathcal{C}_1$ has three components $C_1,C_2,C_3$. 
Suppose $A_1,A_2,A_3$ are valid selections for the component $C_1$; $A_1,A_3$ are valid selections 
for the component $C_2$, and $A_2,A_3$ are valid selections for the component $C_3$.
Clearly, $\rho(A_1)=2$, $\rho(A_2)=2$, $\rho(A_3)=3$, $\rho(A_1A_2)=1$, $\rho(A_1A_3)=2$,
$\rho(A_2A_3)=2$ and $\rho(A_1A_2A_3)=1$. Then $x(A_1),~x(A_2)$ and $x(A_3)$ satisfy the 
following constraints:
\begin{enumerate}
    \item $x(A_1)\leq 2, x(A_2)\leq 2, x(A_3)\leq 3$
    \item $x(A_1)+x(A_2)\leq \rho(A_1) +\rho(A_2)-\rho(A_1A_2)=2+2-1=3$\\
    $x(A_1)+x(A_3)\leq \rho(A_1) +\rho(A_3)-\rho(A_1A_3)=2+3-2=3$\\
    $x(A_2)+x(A_3)\leq \rho(A_2) +\rho(A_3)-\rho(A_2A_3)=2+3-2=3$
    \item $x(A_1)+x(A_2)+x(A_3)=|\mathcal{C}_1|=3$
\end{enumerate}
Therefore the possible solutions are $(x(A_1),x(A_2),x(A_3))=(0,0,3)$, $(0,1,2)$,
$(0,2,1)$, $(1,1,1)$, $(1,0,2)$, $(1,2,0)$,  $(2,0,1)$, $(2,1,0).$ Note that 
$(0,0,3)$ indicates $A_3$ is valid selection in  three components $C_1$,  $C_2$ and $C_3$;
similarly $(0,1,2)$ indicates $A_2$ is a valid selection in one component and $A_3$ is a 
valid selection in two components.

\end{example}
\noindent In the following, we present an ILP formulation for the {\sc Harmless Set} problem for a given $S_X$:\\
      
\noindent       
\vspace{3mm}
    \fbox
    {\begin{minipage}{33.7em}\label{Max-HS}        
\begin{equation*}
\begin{split}
&\text{Maximize} \sum\limits_{l}\sum\limits_{A\in \mathcal{VS}(l)} |A| \times x(A)\\
&\text{Subject to~~~} \\
&x(A_i) \leq \rho(A_i) ~\mbox{ for all }~ A_i\in \mathcal{VS}(l) \\
&x(A_i)+x(A_j) \leq \rho(A_i)+\rho(A_j)-\rho(A_iA_j)~\mbox{ for all }~ A_i,A_j\in \mathcal{VS}(l)\\
&~~~~\vdots\\
&\sum_{A_i\in \mathcal{VS}(l)}{x(A_i)}= |\mathcal{C}_l|\\
&\sum\limits_{l}\sum\limits_{A\in \mathcal{VS}(l)} |N(u) \cap A| \times  x(A) < t(u)~~\mbox{for all }
u\in X 
\end{split}
\end{equation*}  
  \end{minipage} }
    \vspace{3mm}

\noindent In the ILP formulation for {\sc Harmless Set} problem parameterized by vertex integrity, the number of variables are bounded by a computable function $f(k)$. The value of objective function is bounded by $n$ and the value of any variable 
in the integer linear programming is also bounded by $n$. The constraints can be represented using 
$O(k^2 \log{n})$ bits. Proposition \ref{ilp} implies that we can solve the problem
for given $H_X$ in FPT time. 
There are at most $2^k$ guesses for $H_X$, and the ILP formula for a guess can be solved in FPT time. Thus 
Theorem \ref{theorem-vi} holds. 

\section{W[1]-hardness parameterized by treewidth}


In this section  we show that  the {\sc Harmless Set} problem with 
majority thresholds is W[1]-hard when parameterized by  
the treewidth. To show W[1]-hardness of  {\sc Harmless Set} with 
majority thresholds, we reduce from the following problem,  
     which is known to be W[1]-hard parameterized by the treewidth of the graph \cite{DBLP:journals/corr/abs-1107-1177}:
 \vspace{3mm}
    \\
    \fbox
    {\begin{minipage}{33.7em}\label{SP3}
       {\sc  Minimum Maximum Outdegree}\\
        \noindent{\bf Input:} An undirected graph $G$  whose edge weights are given in unary, 
        and a positive integer $r$.\\
    \noindent{\bf Question:} Is there an orientation  of the edges of $G$ such that, for each $v\in V(G)$, 
    the sum of the weights of outgoing edges from $v$ is at most $r$?
    \end{minipage} }\\
    
\noindent In {\sc Minimum Maximum Outdegree} problem, every edge weight $w(u,v)$ of $G$ is given in unary, that is, 
 every edge weight $w(u,v)$ is  polynomially bounded in $|V(G)|$. 
 In a weighted undirected graph $G$, the weighted degree of a vertex $v$, is defined as the sum of the 
 weights of the edges incident to $v$ in $G$.
 In this section, we prove the following theorem:

 \begin{theorem}\label{twtheorem}\rm
 The {\sc Harmless Set} problem with majority thresholds is W[1]-hard when parameterized by the treewidth of the graph.
 \end{theorem}
 
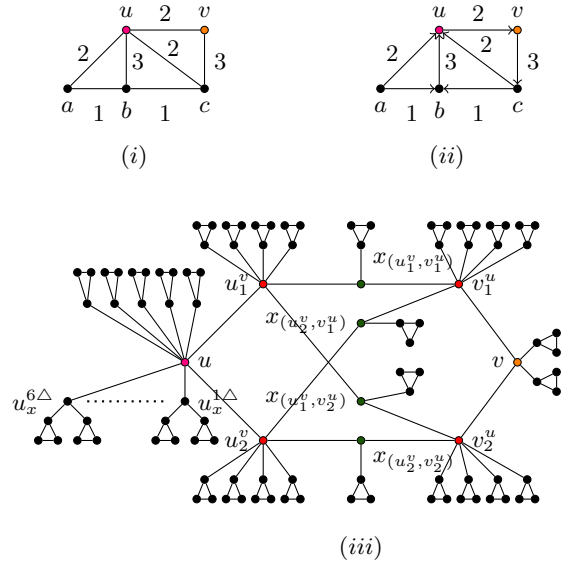
\begin{figure}
\usetikzlibrary{shapes.geometric}
\definecolor{brightpink}{rgb}{1.0, 0.0, 0.5}
    \centering
\begin{tikzpicture}[scale=0.52]

\node[circle,fill=brightpink, draw, inner sep=0 pt, minimum size=0.1cm](u1) at (-2,1.5) [label=above:$u$]{};

\node[circle,fill=orange, draw, inner sep=0 pt, minimum size=0.1cm](v1) at (0,1.5) [label=above:$v$]{};

\node[circle,fill=black, draw, inner sep=0 pt, minimum size=0.1cm](b1) at (-2,0) [label=below:$b$]{};

\node[circle,fill=black, draw, inner sep=0 pt, minimum size=0.1cm](c1) at (0,0) [label=below:$c$]{};

\node[circle,fill=black, draw, inner sep=0 pt, minimum size=0.1cm](a1) at (-3.5,0) [label=below:$a$]{};

\node(x) at (-1,1.3) [label=above:$2$]{};
\node(x) at (-0.8,.4) [label=above:$2$]{};
\node(x) at (-0.2,.7) [label=right:$3$]{};
\node(x) at (-2.3,.7) [label=right:$3$]{};
\node(x) at (-2.5,.9) [label=left:$2$]{};
\node(x) at (-2.7,0) [label=below:$1$]{};
\node(x) at (-1,0) [label=below:$1$]{};

\node(x) at (-1.8,-1) [label=below:$(\romannumeral 1)$]{};
\node(x) at (6.2,-1) [label=below:$(\romannumeral 2)$]{};

\node[circle,fill=brightpink, draw, inner sep=0 pt, minimum size=0.1cm](u2) at (6,1.5) [label=above:$u$]{};

\node[circle,fill=orange, draw, inner sep=0 pt, minimum size=0.1cm](v2) at (8,1.5) [label=above:$v$]{};

\node[circle,fill=black, draw, inner sep=0 pt, minimum size=0.1cm](b2) at (6,0) [label=below:$b$]{};

\node[circle,fill=black, draw, inner sep=0 pt, minimum size=0.1cm](c2) at (8,0) [label=below:$c$]{};

\node[circle,fill=black, draw, inner sep=0 pt, minimum size=0.1cm](a2) at (4.5,0) [label=below:$a$]{};

\node(x) at (7,1.3) [label=above:$2$]{};
\node(x) at (7.2,0.5) [label=above:$2$]{};
\node(x) at (7.8,0.7) [label=right:$3$]{};
\node(x) at (5.7,0.7) [label=right:$3$]{};
\node(x) at (5.4,0.9) [label=left:$2$]{};
\node(x) at (5.3,0) [label=below:$1$]{};
\node(x) at (7,0) [label=below:$1$]{};

\draw(u1)--(v1);
\draw(v1)--(c1);
\draw(c1)--(b1);
\draw(b1)--(a1);
\draw(a1)--(u1);
\draw(u1)--(c1);
\draw(u1)--(b1);

\draw[->](u2)--(v2);
\draw[->](v2)--(c2);
\draw[->](c2)--(b2);
\draw[->](a2)--(b2);
\draw[->](a2)--(u2);
\draw[->](c2)--(u2);
\draw[->](b2)--(u2);

\node[circle,fill=brightpink, draw, inner sep=0 pt, minimum size=0.1cm](u3) at (-0.5,-7) [label=right:$u$]{};

\node[circle,fill=red, draw, inner sep=0 pt, minimum size=0.1cm](u1v) at (1.5,-5) [label=left:$u_{1}^{v}$]{};

\node[circle,fill=red, draw, inner sep=0 pt, minimum size=0.1cm](u2v) at (1.5,-9) [label=left:$u_{2}^{v}$]{};

\node[circle,fill=orange, draw, inner sep=0 pt, minimum size=0.1cm](v3) at (8,-7) [label=left:$v$]{};

\node[circle,fill=red, draw, inner sep=0 pt, minimum size=0.1cm](v1u) at (6.5,-5) [label=right:$v_{1}^{u}$]{};

\node[circle,fill=red, draw, inner sep=0 pt, minimum size=0.1cm](v2u) at (6.5,-9) [label=right:$v_{2}^{u}$]{};

\node[circle,fill=green, draw, inner sep=0 pt, minimum size=0.1cm](xu1vv1u) at (4,-5) [label=above right:$x_{(u_{1}^{v},v_{1}^{u})}$]{};

\node[circle,fill=green, draw, inner sep=0 pt, minimum size=0.1cm](xu2vv1u) at (4,-6) [label= left:$x_{(u_{2}^{v},v_{1}^{u})}$]{};

\node[circle,fill=green, draw, inner sep=0 pt, minimum size=0.1cm](xu2vv2u) at (4,-9) [label= below right:$x_{(u_{2}^{v},v_{2}^{u})}$]{};

\node[circle,fill=green, draw, inner sep=0 pt, minimum size=0.1cm](xu1vv2u) at (4,-8) [label= left: $x_{(u_{1}^{v},v_{2}^{u})}$]{};

\draw(v3)--(v1u);
\draw(v3)--(v2u);
\draw(u3)--(u1v);
\draw(u3)--(u2v);
\draw(u1v)--(xu1vv1u);
\draw(v1u)--(xu1vv1u);
\draw(v1u)--(xu2vv1u);
\draw(u2v)--(xu2vv1u);
\draw(v2u)--(xu2vv2u);
\draw(u2v)--(xu2vv2u);
\draw(u1v)--(xu1vv2u);
\draw(v2u)--(xu1vv2u);


\node[circle,fill=black, draw, inner sep=0 pt, minimum size=0.1cm](x01v3) at (8.5,-6.5) [label=above:]{};

\node[circle,fill=black, draw, inner sep=0 pt, minimum size=0.1cm](x02v3) at (9,-6.25) [label=above:]{};

\node[circle,fill=black, draw, inner sep=0 pt, minimum size=0.1cm](x03v3) at (9,-6.75) [label=above:]{};

\node[circle,fill=black, draw, inner sep=0 pt, minimum size=0.1cm](x11v3) at (8.5,-7.5) [label=above:]{};

\node[circle,fill=black, draw, inner sep=0 pt, minimum size=0.1cm](x12v3) at (9,-7.25) [label=above:]{};

\node[circle,fill=black, draw, inner sep=0 pt, minimum size=0.1cm](x13v3) at (9,-7.75) [label=above:]{};

\draw(v3)--(x01v3);
\draw(x02v3)--(x01v3);
\draw(x03v3)--(x02v3);
\draw(x03v3)--(x01v3);
\draw(v3)--(x11v3);
\draw(x12v3)--(x11v3);
\draw(x13v3)--(x12v3);
\draw(x13v3)--(x11v3);

\node[circle,fill=black, draw, inner sep=0 pt, minimum size=0.1cm](x1u3) at (-3,-5.5) [label=above:]{};
\node[circle,fill=black, draw, inner sep=0 pt, minimum size=0.1cm](x2u3) at (-0.2,-5.5) [label=above:]{};
\node[circle,fill=black, draw, inner sep=0 pt, minimum size=0.1cm](x3u3) at (-0.9,-5.5) [label=above:]{};
\node[circle,fill=black, draw, inner sep=0 pt, minimum size=0.1cm](x4u3) at (-1.6,-5.5) [label=above:]{};
\node[circle,fill=black, draw, inner sep=0 pt, minimum size=0.1cm](x5u3) at (-2.3,-5.5) [label=above:]{};

\node[circle,fill=black, draw, inner sep=0 pt, minimum size=0.1cm](x01u3) at (-2.9,-4.7) [label=above:]{};
\node[circle,fill=black, draw, inner sep=0 pt, minimum size=0.1cm](x02u3) at (-0.1,-4.7) [label=above:]{};
\node[circle,fill=black, draw, inner sep=0 pt, minimum size=0.1cm](x03u3) at (-0.8,-4.7) [label=above:]{};
\node[circle,fill=black, draw, inner sep=0 pt, minimum size=0.1cm](x04u3) at (-1.5,-4.7) [label=above:]{};
\node[circle,fill=black, draw, inner sep=0 pt, minimum size=0.1cm](x05u3) at (-2.2,-4.7) [label=above:]{};

\node[circle,fill=black, draw, inner sep=0 pt, minimum size=0.1cm](x11u3) at (-3.25,-4.7) [label=above:]{};
\node[circle,fill=black, draw, inner sep=0 pt, minimum size=0.1cm](x12u3) at (-0.45,-4.7) [label=above:]{};
\node[circle,fill=black, draw, inner sep=0 pt, minimum size=0.1cm](x13u3) at (-1.15,-4.7) [label=above:]{};
\node[circle,fill=black, draw, inner sep=0 pt, minimum size=0.1cm](x14u3) at (-1.85,-4.7) [label=above:]{};
\node[circle,fill=black, draw, inner sep=0 pt, minimum size=0.1cm](x15u3) at (-2.55,-4.7) [label=above:]{};

\draw(x01u3)--(x11u3);
\draw(x1u3)--(x11u3);
\draw(x1u3)--(x01u3);
\draw(x02u3)--(x12u3);
\draw(x2u3)--(x12u3);
\draw(x2u3)--(x02u3);
\draw(x03u3)--(x13u3);
\draw(x3u3)--(x13u3);
\draw(x3u3)--(x03u3);
\draw(x04u3)--(x14u3);
\draw(x4u3)--(x14u3);
\draw(x4u3)--(x04u3);
\draw(x05u3)--(x15u3);
\draw(x5u3)--(x15u3);
\draw(x5u3)--(x05u3);
\draw(u3)--(x1u3);
\draw(u3)--(x2u3);
\draw(u3)--(x3u3);
\draw(u3)--(x4u3);
\draw(u3)--(x5u3);

\node[circle,fill=black, draw, inner sep=0 pt, minimum size=0.1cm](y01u3) at (-0.5,-8) [label=right:$u^{1 \triangle}_{x}$]{};

\node[circle,fill=black, draw, inner sep=0 pt, minimum size=0.1cm](y11u3) at (0,-8.5) [label=right:]{};

\node[circle,fill=black, draw, inner sep=0 pt, minimum size=0.1cm](y111u3) at (0.25,-9) [label=right:]{};
\node[circle,fill=black, draw, inner sep=0 pt, minimum size=0.1cm](y112u3) at (-0.25,-9) [label=right:]{};

\node[circle,fill=black, draw, inner sep=0 pt, minimum size=0.1cm](y12u3) at (-1,-8.5) [label=right:]{};

\node[circle,fill=black, draw, inner sep=0 pt, minimum size=0.1cm](y121u3) at (-1.25,-9) [label=right:]{};
\node[circle,fill=black, draw, inner sep=0 pt, minimum size=0.1cm](y122u3) at (-0.75,-9) [label=right:]{};

\node[circle,fill=black, draw, inner sep=0 pt, minimum size=0.1cm](z01u3) at (-3.5,-8) [label=left:$u^{6 \triangle}_{x}$]{};

\node[circle,fill=black, draw, inner sep=0 pt, minimum size=0.1cm](z11u3) at (-3,-8.5) [label=right:]{};

\node[circle,fill=black, draw, inner sep=0 pt, minimum size=0.1cm](z111u3) at (-3.25,-9) [label=right:]{};
\node[circle,fill=black, draw, inner sep=0 pt, minimum size=0.1cm](z112u3) at (-2.75,-9) [label=right:]{};

\node[circle,fill=black, draw, inner sep=0 pt, minimum size=0.1cm](z12u3) at (-4,-8.5) [label=right:]{};

\node[circle,fill=black, draw, inner sep=0 pt, minimum size=0.1cm](z121u3) at (-4.25,-9) [label=right:]{};
\node[circle,fill=black, draw, inner sep=0 pt, minimum size=0.1cm](z122u3) at (-3.75,-9) [label=right:]{};

\draw[dotted, thick](-3,-8)--(-1,-8);

\draw(u3)--(z01u3);
\draw(u3)--(y01u3);
\draw(y11u3)--(y01u3);
\draw(y12u3)--(y01u3);
\draw(y11u3)--(y111u3);
\draw(y11u3)--(y112u3);
\draw(y112u3)--(y111u3);
\draw(y12u3)--(y121u3);
\draw(y12u3)--(y122u3);
\draw(y122u3)--(y121u3);
\draw(z11u3)--(z01u3);
\draw(z12u3)--(z01u3);
\draw(z11u3)--(z111u3);
\draw(z11u3)--(z112u3);
\draw(z12u3)--(z121u3);
\draw(z12u3)--(z122u3);
\draw(z112u3)--(z111u3);
\draw(z122u3)--(z121u3);


\node[circle,fill=black, draw, inner sep=0 pt, minimum size=0.1cm](x1u1v) at (0,-4) [label=below:]{};
\node[circle,fill=black, draw, inner sep=0 pt, minimum size=0.1cm](x2u1v) at (0.75,-4) [label=below:]{};
\node[circle,fill=black, draw, inner sep=0 pt, minimum size=0.1cm](x3u1v) at (1.5,-4) [label=below:]{};
\node[circle,fill=black, draw, inner sep=0 pt, minimum size=0.1cm](x4u1v) at (2.25,-4) [label=below:]{};

\node[circle,fill=black, draw, inner sep=0 pt, minimum size=0.1cm](x01u1v) at (-0.2,-3.5) [label=below:]{};
\node[circle,fill=black, draw, inner sep=0 pt, minimum size=0.1cm](x02u1v) at (0.55,-3.5) [label=below:]{};
\node[circle,fill=black, draw, inner sep=0 pt, minimum size=0.1cm](x03u1v) at (1.3,-3.5) [label=below:]{};
\node[circle,fill=black, draw, inner sep=0 pt, minimum size=0.1cm](x04u1v) at (2.05,-3.5) [label=below:]{};

\node[circle,fill=black, draw, inner sep=0 pt, minimum size=0.1cm](x11u1v) at (0.2,-3.5) [label=below:]{};
\node[circle,fill=black, draw, inner sep=0 pt, minimum size=0.1cm](x12u1v) at (0.95,-3.5) [label=below:]{};
\node[circle,fill=black, draw, inner sep=0 pt, minimum size=0.1cm](x13u1v) at (1.7,-3.5) [label=below:]{};
\node[circle,fill=black, draw, inner sep=0 pt, minimum size=0.1cm](x14u1v) at (2.45,-3.5) [label=below:]{};

\draw(u1v)--(x1u1v);
\draw(u1v)--(x2u1v);
\draw(u1v)--(x3u1v);
\draw(u1v)--(x4u1v);

\draw(x01u1v)--(x1u1v);
\draw(x01u1v)--(x11u1v);
\draw(x1u1v)--(x11u1v);

\draw(x02u1v)--(x2u1v);
\draw(x02u1v)--(x12u1v);
\draw(x2u1v)--(x12u1v);

\draw(x03u1v)--(x3u1v);
\draw(x03u1v)--(x13u1v);
\draw(x3u1v)--(x13u1v);

\draw(x04u1v)--(x4u1v);
\draw(x04u1v)--(x14u1v);
\draw(x4u1v)--(x14u1v);

\node[circle,fill=black, draw, inner sep=0 pt, minimum size=0.1cm](0xv1u) at (6,-4) [label=below:]{};
\node[circle,fill=black, draw, inner sep=0 pt, minimum size=0.1cm](1xv1u) at (6.75,-4) [label=below:]{};
\node[circle,fill=black, draw, inner sep=0 pt, minimum size=0.1cm](2xv1u) at (7.5,-4) [label=below:]{};
\node[circle,fill=black, draw, inner sep=0 pt, minimum size=0.1cm](3xv1u) at (8.25,-4) [label=below:]{};

\node[circle,fill=black, draw, inner sep=0 pt, minimum size=0.1cm](x01v1u) at (5.8,-3.5) [label=below:]{};
\node[circle,fill=black, draw, inner sep=0 pt, minimum size=0.1cm](x02v1u) at (6.55,-3.5) [label=below:]{};
\node[circle,fill=black, draw, inner sep=0 pt, minimum size=0.1cm](x03v1u) at (7.3,-3.5) [label=below:]{};
\node[circle,fill=black, draw, inner sep=0 pt, minimum size=0.1cm](x04v1u) at (8.05,-3.5) [label=below:]{};

\node[circle,fill=black, draw, inner sep=0 pt, minimum size=0.1cm](x11v1u) at (6.2,-3.5) [label=below:]{};
\node[circle,fill=black, draw, inner sep=0 pt, minimum size=0.1cm](x12v1u) at (6.95,-3.5) [label=below:]{};
\node[circle,fill=black, draw, inner sep=0 pt, minimum size=0.1cm](x13v1u) at (7.7,-3.5) [label=below:]{};
\node[circle,fill=black, draw, inner sep=0 pt, minimum size=0.1cm](x14v1u) at (8.45,-3.5) [label=below:]{};

\draw(v1u)--(0xv1u);
\draw(v1u)--(1xv1u);
\draw(v1u)--(2xv1u);
\draw(v1u)--(3xv1u);

\draw(x01v1u)--(0xv1u);
\draw(x11v1u)--(0xv1u);
\draw(x01v1u)--(x11v1u);
\draw(x02v1u)--(1xv1u);
\draw(x12v1u)--(1xv1u);
\draw(x02v1u)--(x12v1u);
\draw(x03v1u)--(2xv1u);
\draw(x13v1u)--(2xv1u);
\draw(x03v1u)--(x13v1u);
\draw(x04v1u)--(3xv1u);
\draw(x14v1u)--(3xv1u);
\draw(x04v1u)--(x14v1u);

\node[circle,fill=black, draw, inner sep=0 pt, minimum size=0.1cm](0xu2v) at (0,-10) [label=below:]{};
\node[circle,fill=black, draw, inner sep=0 pt, minimum size=0.1cm](1xu2v) at (0.75,-10) [label=below:]{};
\node[circle,fill=black, draw, inner sep=0 pt, minimum size=0.1cm](2xu2v) at (1.5,-10) [label=below:]{};
\node[circle,fill=black, draw, inner sep=0 pt, minimum size=0.1cm](3xu2v) at (2.25,-10) [label=below:]{};

\node[circle,fill=black, draw, inner sep=0 pt, minimum size=0.1cm](x01u2v) at (-0.2,-10.5) [label=below:]{};
\node[circle,fill=black, draw, inner sep=0 pt, minimum size=0.1cm](x02u2v) at (0.55,-10.5) [label=below:]{};
\node[circle,fill=black, draw, inner sep=0 pt, minimum size=0.1cm](x03u2v) at (1.3,-10.5) [label=below:]{};
\node[circle,fill=black, draw, inner sep=0 pt, minimum size=0.1cm](x04u2v) at (2.05,-10.5) [label=below:]{};

\node[circle,fill=black, draw, inner sep=0 pt, minimum size=0.1cm](x11u2v) at (0.2,-10.5) [label=below:]{};
\node[circle,fill=black, draw, inner sep=0 pt, minimum size=0.1cm](x12u2v) at (0.95,-10.5) [label=below:]{};
\node[circle,fill=black, draw, inner sep=0 pt, minimum size=0.1cm](x13u2v) at (1.7,-10.5) [label=below:]{};
\node[circle,fill=black, draw, inner sep=0 pt, minimum size=0.1cm](x14u2v) at (2.45,-10.5) [label=below:]{};

\draw(u2v)--(0xu2v);
\draw(u2v)--(1xu2v);
\draw(u2v)--(2xu2v);
\draw(u2v)--(3xu2v);

\draw(x01u2v)--(0xu2v);
\draw(x11u2v)--(0xu2v);
\draw(x01u2v)--(x11u2v);
\draw(x02u2v)--(1xu2v);
\draw(x12u2v)--(1xu2v);
\draw(x02u2v)--(x12u2v);
\draw(x03u2v)--(2xu2v);
\draw(x13u2v)--(2xu2v);
\draw(x03u2v)--(x13u2v);
\draw(x04u2v)--(3xu2v);
\draw(x14u2v)--(3xu2v);
\draw(x04u2v)--(x14u2v);

\node[circle,fill=black, draw, inner sep=0 pt, minimum size=0.1cm](0xu2v) at (6,-10) [label=below:]{};
\node[circle,fill=black, draw, inner sep=0 pt, minimum size=0.1cm](1xu2v) at (6.75,-10) [label=below:]{};
\node[circle,fill=black, draw, inner sep=0 pt, minimum size=0.1cm](2xu2v) at (7.5,-10) [label=below:]{};
\node[circle,fill=black, draw, inner sep=0 pt, minimum size=0.1cm](3xu2v) at (8.25,-10) [label=below:]{};

\node[circle,fill=black, draw, inner sep=0 pt, minimum size=0.1cm](x01u2v) at (5.8,-10.5) [label=below:]{};
\node[circle,fill=black, draw, inner sep=0 pt, minimum size=0.1cm](x02u2v) at (6.55,-10.5) [label=below:]{};
\node[circle,fill=black, draw, inner sep=0 pt, minimum size=0.1cm](x03u2v) at (7.3,-10.5) [label=below:]{};
\node[circle,fill=black, draw, inner sep=0 pt, minimum size=0.1cm](x04u2v) at (8.05,-10.5) [label=below:]{};

\node[circle,fill=black, draw, inner sep=0 pt, minimum size=0.1cm](x11u2v) at (6.2,-10.5) [label=below:]{};
\node[circle,fill=black, draw, inner sep=0 pt, minimum size=0.1cm](x12u2v) at (6.95,-10.5) [label=below:]{};
\node[circle,fill=black, draw, inner sep=0 pt, minimum size=0.1cm](x13u2v) at (7.7,-10.5) [label=below:]{};
\node[circle,fill=black, draw, inner sep=0 pt, minimum size=0.1cm](x14u2v) at (8.45,-10.5) [label=below:]{};

\draw(v2u)--(0xu2v);
\draw(v2u)--(1xu2v);
\draw(v2u)--(2xu2v);
\draw(v2u)--(3xu2v);

\draw(x01u2v)--(0xu2v);
\draw(x11u2v)--(0xu2v);
\draw(x01u2v)--(x11u2v);
\draw(x02u2v)--(1xu2v);
\draw(x12u2v)--(1xu2v);
\draw(x02u2v)--(x12u2v);
\draw(x03u2v)--(2xu2v);
\draw(x13u2v)--(2xu2v);
\draw(x03u2v)--(x13u2v);
\draw(x04u2v)--(3xu2v);
\draw(x14u2v)--(3xu2v);
\draw(x04u2v)--(x14u2v);

\node[circle,fill=black, draw, inner sep=0 pt, minimum size=0.1cm](0xu1vv1u) at (4,-4) [label=above right:]{};
\node[circle,fill=black, draw, inner sep=0 pt, minimum size=0.1cm](1xu1vv1u) at (3.75,-3.5) [label=above right:]{};
\node[circle,fill=black, draw, inner sep=0 pt, minimum size=0.1cm](2xu1vv1u) at (4.25,-3.5) [label=above right:]{};

\draw(xu1vv1u)--(0xu1vv1u);
\draw(0xu1vv1u)--(1xu1vv1u);
\draw(0xu1vv1u)--(2xu1vv1u);
\draw(2xu1vv1u)--(1xu1vv1u);

\node[circle,fill=black, draw, inner sep=0 pt, minimum size=0.1cm](0xu2vv2u) at (4,-10) [label=above right:]{};
\node[circle,fill=black, draw, inner sep=0 pt, minimum size=0.1cm](1xu2vv2u) at (3.75,-10.5) [label=above right:]{};
\node[circle,fill=black, draw, inner sep=0 pt, minimum size=0.1cm](2xu2vv2u) at (4.25,-10.5) [label=above right:]{};

\draw(xu2vv2u)--(0xu2vv2u);
\draw(0xu2vv2u)--(1xu2vv2u);
\draw(0xu2vv2u)--(2xu2vv2u);
\draw(2xu2vv2u)--(1xu2vv2u);

\node[circle,fill=black, draw, inner sep=0 pt, minimum size=0.1cm](0xu2vv1u) at (5,-6) [label= right:]{};
\node[circle,fill=black, draw, inner sep=0 pt, minimum size=0.1cm](1xu2vv1u) at (5.5,-6) [label= right:]{};
\node[circle,fill=black, draw, inner sep=0 pt, minimum size=0.1cm](2xu2vv1u) at (5.25,-6.5) [label= right:]{};

\draw(0xu2vv1u)--(xu2vv1u);
\draw(0xu2vv1u)--(1xu2vv1u);
\draw(0xu2vv1u)--(2xu2vv1u);
\draw(2xu2vv1u)--(1xu2vv1u);

\node[circle,fill=black, draw, inner sep=0 pt, minimum size=0.1cm](0xu1vv2u) at (5,-7.25) [label= right:]{};
\node[circle,fill=black, draw, inner sep=0 pt, minimum size=0.1cm](1xu1vv2u) at (5.5,-7.25) [label= right:]{};
\node[circle,fill=black, draw, inner sep=0 pt, minimum size=0.1cm](2xu1vv2u) at (5.25,-7.75) [label= right:]{};

\draw(0xu1vv2u)--(1xu1vv2u);
\draw(0xu1vv2u)--(2xu1vv2u);
\draw(2xu1vv2u)--(1xu1vv2u);
\draw(xu1vv2u)--(2xu1vv2u);

\node(x) at (4,-11) [label=below:$(\romannumeral 3)$]{};

\end{tikzpicture}
    \caption{$(\romannumeral1)$ An instance $(G,r)$ of Minimum Maximum Outdegree with $r=3$. $(\romannumeral 2)$ A valid orientation of $G$ when $r=3$. $(\romannumeral 3)$ An illustration of the reduction algorithm in Theorem \ref{twtheorem} using an edge $(u,v)$ with $\omega(u,v)=2$. Note that $u$ is a {\it high-degree-type} vertex and 
    $v$ is a {\it low-degree-type vertex}.}
    \label{edgedel1}
\end{figure}

\proof Let  $G=(V,E,w)$ and a positive integer $r \geq 3$ be an  instance $I$ of 
{\sc Minimum Maximum Outdegree}. We construct an instance $I'=(G',t,k)$ 
of {\sc Harmless Set} the following way. See Figure \ref{edgedel1} for an illustration.
\begin{enumerate}
    \item 

For each weighted edge 
$(u,v)\in E(G)$, we introduce two sets of new vertices $V_{uv}=\{u^v_1, \ldots, u^v_{w(u,v)}\}$ and
$V_{vu}=\{v^u_1, \ldots, v^u_{w(u,v)}\}$ into $G'$. 
 Make $u$ and $v$ adjacent to every vertex of $V_{uv}$ and
$V_{vu}$, respectively. The vertices of $\bigcup\limits_{(u,v)\in E(G)}{}{V_{uv}\cup V_{vu}}$ are called 
{\it type 1} vertices. 
\item  For each $1 \leq i \leq w(u,v)-1$, we introduce 
 $x_{(u^v_i,v^u_i)}$ into $G'$ and make it adjacent to $u^{v}_{i}$ and $v^{u}_{i}$;
 introduce 
 $x_{(u^v_i,v^u_{i+1})}$  and make it adjacent to $u^{v}_{i}$ and $v^{u}_{i+1}$;
 introduce 
 $x_{(u^v_{i+1},v^u_i)}$ and make it adjacent to $u^{v}_{i+1}$ and $v^{u}_{i}$. 
 We also add $x_{(u^v_{w(u,v)},v^u_{w(u,v)})}$ into $G'$ and make it 
 adjacent to $u^v_{w(u,v)}$ and $v^u_{w(u,v)}$.  We call such vertices, the vertices of {\it type 2}. 
 \item For every vertex $x$ of {\it type 2}, we add a triangle (cycle of length 3) and make $x$ adjacent to exactly one vertex of this triangle. For every vertex $x$ of {\it type 1},  let  $n(x)$ be the number of neighbours of $x$ in  $V(G)$ and 
 in the set of {\it type 2} vertices.
  Note that $2 \leq n(x)\leq 4$. We add $n(x)+1$ many triangles corresponds to vertex $x$ and make $x$ adjacent to exactly one  vertex of each triagle. 
  \item The weighted degree of a vertex $x\in V$ in $G$ is denoted by  $d_{w}(x;G)$. 
 We partition the vertices of $V(G)$ based on whether 
 $ \lceil \frac {d_{w}(x;G)}{2} \rceil \leq r+1$ or 
 $ \lceil \frac {d_{w}(x;G)}{2} \rceil > r+1$. 
 A vertex $x$ in $G$ with $ \lceil \frac {d_{w}(x;G)}{2} \rceil \leq r+1$
 is called a vertex of {\it low-degree-type}.
 For each $x \in V(G)$ of {\it low-degree-type}, 
 we add  
 $2\Big[(r+1) - \lceil \frac {d_{w}(x;G)}{2} \rceil \Big]$  triangles and make $x$ adjacent to exactly one vertex of each triangle. 
 A vertex $x \in V(G)$ with $ \lceil \frac {d_{w}(x;G)}{2} \rceil > r+1$
 is called a vertex of {\it high-degree-type}.
 For each  $x\in V(G)$ of {\it high-degree-type},  
 we add a set $V_{x}^{\triangle}=\{v_{x}^{1 \triangle}, \ldots, v_{x}^{ \alpha \triangle} \}$ of $\alpha=d_{w}(x;G)-r$ many vertices and 
 make them adjacent to $x$. 
  For each $v \in V_{x}^{\triangle}$, we add two triangles and make $v$ adjacent to exactly one  vertex of each triangle.
  For each {\it high-degree-type} vertex $x$, we also add a set  of $(r+2) $ many triangles and make $x$ adjacent 
 to exactly one  vertex of each triangle.
 \end{enumerate}
  This completes the construction of graph $G'$.

\noindent We set $k = n+ W + \sum\limits_{(u,v)\in E(G)} (3w(u,v)-2) + \sum\limits_{x \in \text{\it high-degree-type}} (d_w(x;G)-r)$ where $W=\sum\limits_{(u,v)\in E(G)}{w(u,v)}$. 
 Clearly $I'$ can be computed 
 in polynomial time.  We now show that  the treewidth of $G'$  is bounded by a function of  the treewidth of $G$. 
 We do so by modifying an optimal tree decomposition $T$ of $G$ as follows:
 \begin{itemize}
     \item For every edge $(u,v)$ of $G$, we take an arbitrary  node $t$ in $T$ whose  bag $X_t$  contains 
     both $u$ and $v$; attach to this node a chain of nodes $1,2,\ldots,w(u,v)-1$ such that the bag of node $i$
     is $$X_t\cup \{u_i^v, v_{i}^{u}, u_{i+1}^{ v},
     v_{i+1}^{ u}, x_{(u_i^v, v_{i}^{u})},x_{( u_{i+1}^{ v},
     v_{i+1}^{ u})},x_{(u_i^v,v_{i+1}^{ u})},x_{(v_{i}^{u},u_{i+1}^{ v})}\}.$$ 
     \item For every {\it Type 1} vertex $u^{v}$, take an arbitrary  node $t$  in the modified
     tree decomposition whose  bag 
     contains $u^{v}$; attach to it a chain of at most five nodes $t_1,t_2,\ldots,t_5$  such that 
     the bag $X_{t_i}$ of node $t_i$ 
     contains $u^v$ and the vertices of $i$th triangle corresponds to $u^{v}$. 
    \item For every {\it Type 2} vertex $x_{(u^{v},v^{u})}$, take an arbitrary node 
    $t$ in the modified
     tree decomposition whose  bag 
     contains  $x_{(u^{v},v^{u})}$; attach to it another node $t'$  such that 
     the bag $X_{t'}$ of node $t'$ 
     contains $x_{(u^{v},v^{u})}$ and the vertices of 
     the triangle corresponds to $x_{(u^{v},v^{u})}$. 
     
     \item For every edge $(u,v)$ of $G$, we take an arbitrary node $t$ in $T$ whose  bag $X_t$  contains $u$. If $u$ is of {\it low-degree-type} then attach to it a 
     chain of $2\big[(r+1)-\lceil \frac{d_{w}(x;G)}{2} \rceil \big]$ nodes such that 
     the bag of node $i$ contains  $u$ and the vertices of $i$th triangle corresponds to $u$.

    \item  For every edge $(u,v)$ of $G$, we take an arbitrary node $t$ in $T$ whose  bag $X_t$  contains $u$. If $u$ is of {\it high-degree-type} then attach to it two 
    chains of node:  the first chain of node $1,2,\ldots, r+2$  such that the bag 
    $X_{i}$ of node $i$ contains  $u$ and the vertices of $i$ triangle
    corresponds to $u$; and the second chain of nodes  $1,2,\ldots, d_{w}(x;G)-r$ 
    such that the bag $X_i$ of node $i$ contains
    $u,v_{x}^{i \triangle}$ and the vertices of two triangles corresponds to $v_{x}^{i \triangle}$.

 \end{itemize}
 It is easy to verify that the result is a valid tree decomposition of  $G'$ and 
 its width is at most the treewidth of $G$ plus eight.

\par Now we show that our reduction is correct. That is, we prove that $I$ is  a yes instance of 
{\sc Minimum Maximum Outdegree} if and only if $I^{\prime}$ is a yes instance of {\sc Harmless Set}.
Let $D$ be the directed graph obtained by  an orientation of the edges of $G$ such that for each vertex the sum of the weights of outgoing edges is at most $r$. We claim that the set
\begin{equation*}
\begin{split}
H & = V(G) \bigcup\limits_{(u,v)\in E(D)} V_{uv} \bigcup\limits_{x\in \text{\it high-degree-type}} V_{x}^{\triangle} \\
&\bigcup\limits_{(u,v)\in E(G)} \Big\{x_{(u^v_i,v^u_i)}, x_{(u^v_i,v^u_{i+1})}, x_{(u^v_{i+1},v^u_i)}, x_{(u^v_{\omega(u,v)},v^u_{\omega(u,v)})} ~|~ 1\leq i \leq \omega(u,v)-1 \Big\}
\end{split}          
\end{equation*}
is harmless set of size at least $k$. 
Next, we show that all the vertices in $H$ satisfy the threshold condition. It is easy to verify that  each $u\in \bigcup\limits_{(u,v)\in E(G)} (V_{uv} \cup V_{vu})$
satisfies the threshold condition as $u$ has $n(u)$ neighbours in $H$ and $n(u)+1$ neighbours outside $H$, that is, $u$ has less than $\lceil\frac{d_{G'}(u)}{2}\rceil$ neighbours in $H$. Each  $$x\in \bigcup\limits_{(u,v)\in E(G)} \Big\{x_{(u^v_i,v^u_i)}, x_{(u^v_i,v^u_{i+1})}, x_{(u^v_{i+1},v^u_i)}, x_{(u^v_{w(u,v)},v^u_{w(u,v)})} ~|~ 1\leq i \leq w(u,v)-1 \Big\} $$ satisfies the threshold condition as $x$ has only one neighbour in $H$ and two neighbours outside $H$, that is, 
$x$ has less than $\lceil\frac{d_{G'}(x)}{2}\rceil=\lceil\frac{3}{2}\rceil=2$ neighbours in $H$.  It is also easy to see that the vertices of triangles satisfy the threshold condition. Let $u$ be an arbitrary vertex of {\it low-degree-type}. If the weighted degree of $u$ in $G$ is $d_{w}(u;G)$ then its degree in $G'$ is 
$d_{w}(u;G) + 2\Big[(r+1) - \lceil \frac {d_{w}(u;G)}{2} \rceil \Big]$. 
Observe that the neighbours of $u$ inside $H$ are all of {\it type 1} which is equal to outdegree of $u$ and we know outdegree of $u$ is bounded by $r$. 
Thus each {\it low-degree-type} vertex has less than $\lceil\frac{d_{G'}(u)}{2}\rceil=r+1$ neighbours in $H$.   Therefore each {\it low-degree-type} vertex satisfies the threshold condition. 
Next, let $x$ be an arbitrary vertex of {\it high-degree-type}.  If the weighted degree of $x$ in $G$ is $d_{w}(x;G)$ then its degree in $G'$ is $2d_{w}(x;G)+2 $. Clearly the neighbours of $x$ inside $H$ are at most $r+ (d_{w}(x;G)-r) $. Therefore the  vertices of {\it high-degree-type} satisfy the threshold condition. This implies that $I'$ is a yes instance. 
\par Conversely, assume that $G'$ admits a harmless set $H$ of size at least $k$. 
We make the following observations: (i) let $C$ be the set of all triangles 
introduced in the reduction algorithm, then $C$ does not intersect with $H$. This is true because any vertex with degree 2 has threshold equal to 1. This implies that both the neighbours of that vertex have to be outside the solution as otherwise the vertex will fail to satisfy the threshold condition, (ii) for each $(u,v)\in E(G)$ the set $V_{uv} \cup V_{vu}$ contributes at most half vertices in $H$ as otherwise $x_{(u^v_i,v^u_i)}$ for some 
$1 \leq i \leq w(u,v)$ will fail to satisfy the threshold condition. Note that 
the total number of vertices in $\bigcup\limits_{(u,v)\in E(G)}{V_{uv}\cup V_{vu}}$ is $2W$.
The above observations imply that the size of harmless set $H$ is at most 
$|V(G')|- W- 3|C|= n+ W + \sum\limits_{(u,v)\in E(G)} (3w(u,v)-2) + \sum\limits_{x \in \text{\it high-degree-type}} (d_{w}(x;G)-r)$, which is equal to $k$. It implies that either $V_{uv}\subseteq H$ or $V_{vu}\subseteq H$ for all $(u,v)\in E(G)$ as otherwise some vertex
$x_{(u^v_i,v^u_{i+1})}$ will fail to satisfy the threshold condition. Hence the harmless set  is of the form 
\begin{equation*}
\begin{split}
H=& V(G) \bigcup\limits_{(u,v)\in E(G)} (V_{uv} \ \text{or} \ V_{vu})\bigcup\limits_{x\in \text{\it high-degree-type}} V_{x}^{\triangle} \\
& \bigcup\limits_{(u,v)\in E(G)} \Big\{x_{(u^v_i,v^u_i)}, x_{(u^v_i,v^u_{i+1})}, x_{(u^v_{i+1},v^u_i)}, x_{(u^v_{w(u,v)},v^u_{w(u,v)})} ~|~ 1\leq i \leq w(u,v)-1 \Big\}. 
 \end{split}
 \end{equation*}
 Next, we define a directed graph $D$ by 
$V(D)=V(G)$ and
\[E(D)=\Big\{ (u,v) ~|~u,v\in V(D) \mbox{ and } V_{uv} \subseteq H\Big\}
\bigcup \Big\{ (v,u) ~|~u,v\in V(D) \mbox{ and } V_{vu} \subseteq H\Big\}. \]  Let us assume that there exists a vertex $x\in V(G)$ of {\it low-degree-type} such that the outdegree is more than $r$. We can easily see that $d_{H}(x) \geq \lceil \frac{d_{G'}(x)}{2} \rceil$ which is a contradiction. Let us assume that there exists a vertex $x\in V(G)$ of {\it high-degree-type} such that the outdegree is more than $r$. We can easily see that $d_{H}(x) \geq d_{w}(x;G) \geq \lceil \frac{d_{G'}(x)}{2} \rceil $ which is a contradiction. This implies that $I$ is a 
yes-instance.  \qed \\

\section{Hardness Results}
In this section we show that {\sc Harmless Set}
is W[1]-hard parameterized by a vertex deletion set to trees of height 
at most three, that is, a subset $D$ of the vertices of the graph such that 
every component in the graph, after removing $D$, is a tree of height at 
most three. We show our hardness result for {\sc Harmless Set} using a reduction 
from the {\sc Multidimensional Relaxed Subset Sum (MRSS)} problem. 

\noindent\vspace{3mm}
    \\
    \fbox
    {\begin{minipage}{33.7em}\label{SP1}
       {\sc  Multidimensional Subset Sum (MSS)}\\
     \noindent{\bf  Input:} An integer $k$, a set 
     $S = \{s_1,\ldots,s_n\}$ of vectors with $s_i \in \mathbb{N}^k$ for every $i$ with 
     $1 \leq i \leq  n$  and a target vector $g \in \mathbb{N}^k$.\\
\noindent {\bf Parameter}: $k$ \\
\noindent{\bf Question}: Is there a subset $S'\subseteq S $ such that $\sum\limits_{s\in S'}{s}=g$?
    \end{minipage} }\\

  \noindent  We consider a variant of MSS that we require in our proofs. 
  In the {\sc Multidimensional Relaxed Subset Sum} ({\sc MRSS}) problem, an additional integer $k'$ is given
    (which will be part of the parameter)
    and we ask whether there is a subset $S'\subseteq S$ with $|S'|\leq k'$ such that $\sum\limits_{s\in S'}{s}\geq g$.
    This variant can be formalized as
follows:

\noindent\vspace{3mm}
    \\
    \fbox
    {\begin{minipage}{33.7em}\label{SP2}
       {\sc Multidimensional Relaxed Subset Sum (MRSS)}\\
     \noindent{\bf  Input:} An integer $k$, a set 
     $S = \{s_1,\ldots,s_n\}$ of vectors with $s_i \in \mathbb{N}^k$ for every $i$ with 
     $1 \leq i \leq  n$, a target vector $t \in \mathbb{N}^k$ and an integer $k'$.\\
\noindent {\bf Parameter}: $k+k'$ \\
\noindent{\bf Question}: Is there a subset $S'\subseteq S $ with $|S'|\leq k'$ such that $\sum\limits_{s\in S'}{s}\geq g$?
    \end{minipage} }\\
    
\noindent It is known that {\sc MRSS} is W[1]-hard when parameterized by the combined parameter $k+k'$,
   even if all integers in the input are given in unary \cite{mss}.
 We now prove the following theorem:
 \begin{theorem}\label{trtheorem-gen}\rm
 The {\sc Harmless Set} problem with general thresholds is W[1]-hard when 
 parameterized by the size of a vertex deletion set into
trees of height at most 3, even when restricted to bipartite graphs.
 \end{theorem}

\proof Let  $(k, k', S, g)$  be an instance of {\sc MRSS}.  From this we construct an  instance 
 $(G,t,r)$ of {\sc Harmless Set} the following way. For each vector
$s\in S$, we introduce a tree $T^s$ of height three. 
For $s\in S$, $ \max(s)$ is the value of the largest coordinate of $s$ and 
$\max(S)$ is maximum of $\max(s)$ values. 
The tree $T^s$ consists of vertices $V(T^s) = A^s \cup B^s \cup \{c^s\}$ where $A^s=\{a^s_{1},\ldots,a^s_{\max(S)}\}$ and $B^s=\{b^s_{1},\ldots,b^s_{\max(S)}\}$. 
Make $c^s$  adjacent to every vertex of $B^s$. We also make $a^s_i$ adjacent to 
$b^s_i$ for all $1\leq i \leq \max(s)$.
 Next, we introduce the set $U=\{u_{1},\ldots,u_{k}\}$ of vertices into $G$. 
  For each $1\leq i \leq k$ and $s\in S$, 
  make $u_i$ adjacent to exactly $s(i)$ many vertices of $A^s$ arbitrarily. 
 We introduce three cycles $C_1,C_2,C_3$ of length four where 
 $V(C_1)=\{a_1,a_2,a_3,a_4\}$, $V(C_2)=\{b_1,b_2,b_3,b_4\}$ and 
 $V(C_3)=\{c_1,c_2,c_3,c_4\}$.
 We make $a_1$ adjacent to every vertex of $\bigcup\limits_{s\in S} A^s $,
 make $b_1$ adjacent to every vertex of $\bigcup\limits_{s\in S} B^s$ and 
 make  $c_1$  adjacent to  every vertex of $\bigcup\limits_{s\in S} \{c^s\}$.  This completes the construction of graph $G$. 
  \begin{figure}
    \centering
\begin{tikzpicture}[scale=0.8]

\node[circle,fill=white, draw, inner sep=0 pt, minimum size=0.2cm](cs2) at (0,0) [label=left:$c^{s_{2}}$]{};

\node[circle,fill=orange,draw, inner sep=0 pt, minimum size=0.2cm](b2s2) at (0.75,1) [label=left:$b_{2}^{s_{2}}$]{};
\node[circle,fill=orange,draw, inner sep=0 pt, minimum size=0.2cm](b1s2) at (-0.75,1) [label=left:$b_{1}^{s_{2}}$]{};

\node[circle,fill=orange,draw, inner sep=0 pt, minimum size=0.2cm](a2s2) at (0.75,2) [label=left:$a_{1}^{s_{2}}$]{};
\node[circle,fill=orange,draw, inner sep=0 pt, minimum size=0.2cm](a1s2) at (-0.75,2) [label=left:$a_{2}^{s_{2}}$]{};

\node[circle,fill=orange,draw, inner sep=0 pt, minimum size=0.2cm](cs3) at (3.75,0) [label=right:$c^{s_{3}}$]{};

\node[circle,fill=orange,draw, inner sep=0 pt, minimum size=0.2cm](b2s3) at (4.5,1) [label=left:$b_{2}^{s_{3}}$]{};
\node[circle,fill=orange,draw, inner sep=0 pt, minimum size=0.2cm](b1s3) at (3,1) [label=left:$b_{1}^{s_{3}}$]{};

\node[circle,fill=white,draw, inner sep=0 pt, minimum size=0.2cm](a2s3) at (4.5,2) [label=left:$a_{2}^{s_{3}}$]{};
\node[circle,fill=white,draw, inner sep=0 pt, minimum size=0.2cm](a1s3) at (3,2) [label=left:$a_{1}^{s_{3}}$]{};

\node[circle,fill=orange, draw, inner sep=0 pt, minimum size=0.2cm](cs1) at (-3.75,0) [label=left:$c^{s_{1}}$]{};

\node[circle,fill=orange,draw, inner sep=0 pt, minimum size=0.2cm](b2s1) at (-3,1) [label=left:$b_{2}^{s_{1}}$]{};
\node[circle,fill=orange,draw, inner sep=0 pt, minimum size=0.2cm](b1s1) at (-4.5,1) [label=left:$b_{1}^{s_{1}}$]{};

\node[circle,fill=white,draw, inner sep=0 pt, minimum size=0.2cm](a1s1) at (-4.5,2) [label=left:$a_{1}^{s_{1}}$]{};
\node[circle,fill=white,draw, inner sep=0 pt, minimum size=0.2cm](a2s1) at (-3,2) [label=left:$a_{2}^{s_{1}}$]{};

\node[circle,fill=white, draw, inner sep=0 pt, minimum size=0.2cm](a1) at (0,3.5) [label=left:$a_{1}$]{};

\node[circle,fill=white, draw, inner sep=0 pt, minimum size=0.2cm](a3) at (0,4.5) [label=above:$a_{3}$]{};
\node[circle,fill=white, draw, inner sep=0 pt, minimum size=0.2cm](a2) at (-0.7,4) [label=left:$a_{2}$]{};
\node[circle,fill=white, draw, inner sep=0 pt, minimum size=0.2cm](a4) at (0.7,4) [label=right:$a_{4}$]{};

\node(g) at (0,5) [label=above:$C_{1}$]{};

\node[circle,fill=orange, draw, inner sep=0 pt, minimum size=0.2cm](u1) at (-2,3.5) [label=left:$u_{1}$]{};

\node[circle,fill=orange, draw, inner sep=0 pt, minimum size=0.2cm](u2) at (2,3.5) [label=left:$u_{2}$]{};

\node[circle,fill=white, draw, inner sep=0 pt, minimum size=0.2cm](c1) at (1.3,-1.5) [label=left:$c_{1}$]{};
\node[circle,fill=white, draw, inner sep=0 pt, minimum size=0.2cm](c3) at (1.3,-2.5) [label=below:$c_{3}$]{};
\node[circle,fill=white, draw, inner sep=0 pt, minimum size=0.2cm](c2) at (1.9,-2) [label=right:$c_{4}$]{};
\node[circle,fill=white, draw, inner sep=0 pt, minimum size=0.2cm](c4) at (0.77,-2) [label=left:$c_{2}$]{};

\node[circle,fill=white, draw, inner sep=0 pt, minimum size=0.2cm](b1) at (-1.8,-1.5) [label=left:$b_{1}$]{};
\node[circle,fill=white, draw, inner sep=0 pt, minimum size=0.2cm](b3) at (-1.8,-2.5) [label=below:$b_{3}$]{};
\node[circle,fill=white, draw, inner sep=0 pt, minimum size=0.2cm](b2) at (-2.4,-2) [label=left:$b_{2}$]{};
\node[circle,fill=white, draw, inner sep=0 pt, minimum size=0.2cm](b4) at (-1.2,-2) [label=right:$b_{4}$]{};

\node(g) at (-1.8,-3) [label=below:$C_{2}$]{};

\node(g) at (1.3,-3) [label=below:$C_{3}$]{};

\draw(a1)--(a2);
\draw(a2)--(a3);
\draw(a3)--(a4);
\draw(a4)--(a1);

\draw(b1)--(b2);
\draw(b2)--(b3);
\draw(b3)--(b4);
\draw(b4)--(b1);

\draw(c1)--(c2);
\draw(c2)--(c3);
\draw(c3)--(c4);
\draw(c4)--(c1);

\draw[green](c1)--(cs1);
\draw[green](c1)--(cs2);
\draw[green](c1)--(cs3);

\draw[green](b1)--(b1s1);
\draw[green](b1)--(b2s1);
\draw[green](b1)--(b1s2);
\draw[green](b1)--(b2s2);
\draw[green](b1)--(b1s3);
\draw[green](b1)--(b2s3);

\draw(cs1)--(b1s1);
\draw(cs1)--(b2s1);
\draw(a1s1)--(b1s1);
\draw(a2s1)--(b2s1);

\draw(cs2)--(b1s2);
\draw(cs2)--(b2s2);
\draw(a1s2)--(b1s2);
\draw(a2s2)--(b2s2);

\draw(cs3)--(b1s3);
\draw(cs3)--(b2s3);
\draw(a1s3)--(b1s3);
\draw(a2s3)--(b2s3);

\draw[green](u1)--(a1s1);
\draw[green](u1)--(a2s1);
\draw[green](u1)--(a1s2);
\draw[green](u1)--(a1s3);

\draw[green](u2)--(a2s1);
\draw[green](u2)--(a2s2);
\draw[green](u2)--(a1s3);
\draw[green](u2)--(a2s3);

\draw[green](a1)--(a1s1);
\draw[green](a1)--(a2s1);
\draw[green](a1)--(a1s2);
\draw[green](a1)--(a2s2);
\draw[green](a1)--(a1s3);
\draw[green](a1)--(a2s3);

\end{tikzpicture}
    \caption{The graph $G$ in the proof of Theorem \ref{trtheorem-gen} constructed from MRSS instance $S=\{ (2,1),(1,1),(1,2)\}, g=(3,3), k=2, k'=2$. The set $S'=\{(2,1),(1,2)\}$ forms a solution of MRSS instance and the set
    $H=\{u_1,u_2\} \cup B^{s_1}\cup B^{s_2} \cup B^{s_3}\cup A^{s_2} \cup \{c^{s_1}, c^{s_3}\}  $ forms a harmless set in $G$. }
    \label{thorem5}
\end{figure}
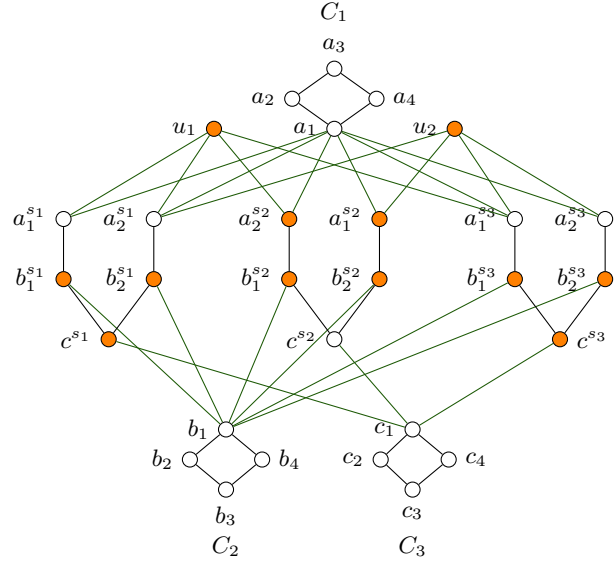
Note that $G$ is a bipartite graph with bipartition 
$$V_{1} = U \bigcup\limits_{s\in S} B^s \bigcup \{a_1,a_3,b_2,b_4, c_1,c_3\}  $$ and $$V_{2} = \bigcup\limits_{s\in S} A^s \bigcup\limits_{s\in S} \{c^s\} \bigcup \{a_2,a_4,b_1,b_3,c_2,c_4\}.$$ We observe that if we delete the set $U \cup \{a_1,a_2,a_3,a_4,b_1,b_2,b_3,b_4,c_1\}$ of size $k+9$ from $G$ 
then we are left with trees of height at most three. We define the threshold function 
as follows:

 \[ t(u) = 
            \begin{cases}
               d(u) &\quad\text{if $u \in \{b_1\} \bigcup\limits_{s\in S} A^s\cup \{c^s\}$}\\
                1 &\quad\text{if $u \in \{a_2,a_3,a_4,b_2,b_3,b_4,c_2,c_3,c_4\}$}\\
               2 & \quad\text{if $u\in \bigcup\limits_{s\in S} B^s $ }\\
               (n-k') \max(S)+1 & \quad\text{if $u=a_1$} \\
               k'+1 & \quad\text{if $u=c_1$} \\
               \sum\limits_{s\in S} s(i) - g(i)+1 & \quad\text{if $u= u_{i}$ for all $1\leq i \leq k$}
            \end{cases} 
            \]
We set $r= k+ n\times \max(S)+ (n-k')\times \max(S) +k'$. Now we show that our reduction is correct. That is, we claim $(k, k', S, g)$ is  a yes instance of 
{\sc MRSS} if and only if $(G,t,r)$ is a yes instance of {\sc Harmless Set}. 
Towards showing the forward direction, let $S'\subseteq S$ be such that 
$|S'|\leq k'$ and $\sum\limits_{s\in S'}{s}\geq g$. We claim that the set 
$$H = U \cup \bigcup\limits_{s\in S} B^s \bigcup\limits_{s\in S\setminus S'} A^s \bigcup\limits_{s\in S'} \{c^s\}$$ is a harmless set of size at least $r$. 
It is easy to see that $|H|\geq r$. Next, we show that all the vertices in $G$ satisfy 
the threshold condition. Let $u_{i}$ be an arbitrary  element in  $U$. 
We know that $d(u_{i})= \sum\limits_{s\in S}s(i)$ and $\sum\limits_{s\in S'}{s}\geq g$. Hence, we get $d_H(u_i)\leq \sum\limits_{s\in S}s(i)-g(i)<t(u_i)=\sum\limits_{s\in S} s(i) - g(i)+1$. 
It is easy to see that every vertex $u$ of  $\bigcup\limits_{s\in S} A^s$ satisfies the threshold condition.  As $a_1\not\in H$, $d_H(u)=d(u)-1< d(u)=t(u)$. Similarly, every vertex of  $\bigcup\limits_{s\in S} \{c_{s}\}$ satisfies the threshold condition as $c_1\not\in H$. 
For each $b^s_i \in \bigcup\limits_{s\in S} B^s$, we have either $a^s_i$ or $c^s$ inside the solution and $b_1\not\in H$. Thus $d_H(b^s_i)=1<2=t(b^s_i)$. Hence all the vertices in 
$\bigcup\limits_{s\in S} B^s$ satisfy the threshold condition. 
As $|S'|\leq k'$, we see that $c_1$ has at most $k'$ neighbours inside $H$, thus $d_H(c_1)=k'<k'+1=t(c_1)$. 
For the rest of the vertices in $H$ it is easy to verify that the threshold condition is satisfied.

\par Towards showing the reverse direction of the claim, let $H$ be a harmless 
set of size at least $r$ in $G$. It is easy to see that 
$H \cap \{a_i,b_i,c_i~|~1\leq i\leq 4\} = \emptyset$ as otherwise one of the 
vertex in $ \{a_i,b_i,c_i~|~1\leq i\leq 4\}$  will fail to satisfy the threshold condition. Clearly $U$ and $\bigcup\limits_{s\in S} B^{s}$ can contribute at most $k$ and $n\times \max(S)$ to the solution respectively. We also observe that the set $\bigcup\limits_{s\in S} A^s$ can contribute at most $(n-k') \times \max(S)$ to the solution due to 
the fact that 
$t(a_1)=(n-k')\times \max(S)+1$. 
Therefore, the only way to have a harmless set of size at least $r$ is that the set $\bigcup\limits_{s\in S} \{c^s\}$ contribute at least $k'$ elements to set $H$. 
Since $t(c_1)=k'+1$, the set $\bigcup\limits_{s\in S} \{c^{s}\}$ contributes 
at most $k'$ elements to set $H$. Therefore, the set $\bigcup\limits_{s\in S} \{c^{s}\}$
contributes exactly $k'$ elements to $H$. We define
$$ S'=\{s\in S~|~c^s\in H\}. $$
Observe that $\bigcup\limits_{s\in S'} A^s \cap H=\emptyset$ as otherwise one of the vertices in the set $\bigcup\limits_{s\in S'} B^s$ will not satisfy the threshold condition. From here, we see that $H= U\cup \bigcup\limits_{s\in S} B^s \bigcup\limits_{s\in S\setminus S'} A^s \bigcup\limits_{s\in S'} \{c^s\}$. Since  each $u_i\in U$ satisfies the 
threshold condition, we have $d_H(u_i)= \sum\limits_{s\in S\setminus S'}{s(i)}=
\sum\limits_{s\in S}{s(i)} -\sum\limits_{s\in S'}{s(i)}<t(u_i)=
\sum\limits_{s\in S} s(i)-g(i)+1$. This implies that $\sum\limits_{s\in S'}s(i)\geq g(i)$ for all $1\leq i\leq k$. Therefore  $(k, k', S, g)$ is a yes-instance. \qed \\

 \noindent Clearly trees of height at most three are trivially acyclic. 
 Moreover, it is easy to verify that such trees have 
 pathwidth \cite{Kloks94} and treedepth \cite{Sparsity} at most three, which implies:
 
\begin{theorem}\rm
 The {\sc Harmless Set} problem with general thresholds 
 is W[1]-hard when parameterized by any of the following parameters:
 \begin{itemize}
     \item the feedback vertex set number,
     \item the pathwidth and treedepth of the input graph,
     \item the size of a minimum set of vertices whose deletion results in components of pathwidth/treedepth at most three,
 \end{itemize}
  even when restricted to bipartite graphs.
\end{theorem}
\section{W[1]-hardness parameterized by cluster vertex deletion number}
The \emph{cluster vertex deletion number} of a graph is the minimum number of its vertices whose deletion results in a disjoint union of complete graphs. This generalizes the vertex cover number, provides an upper bound to the clique-width and is related to the previously studied notion of the twin cover of the graph under consideration. 
 \begin{theorem}\label{trtheorem-cvd}
 The {\sc Harmless Set} problem with general thresholds is W[1]-hard when 
 parameterized by the cluster vertex deletion number of the input graph.
 \end{theorem}
 
\proof Let  $(k, k', S, g)$  be an instance of {\sc MRSS}.  From this we construct an  instance 
 $(G,t,r)$ of {\sc Harmless Set} the following way.
 
   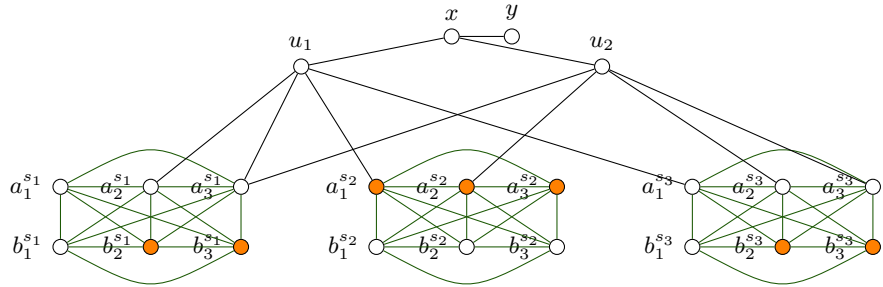
\begin{figure}
    \centering
\begin{tikzpicture}[scale=0.8]

\node[circle,draw, inner sep=0 pt, minimum size=0.2cm](b3s2) at (0.75,1) [label=left:$b_{3}^{s_{2}}$]{};
\node[circle,draw, inner sep=0 pt, minimum size=0.2cm](b2s2) at (-0.75,1) [label=left:$b_{2}^{s_{2}}$]{};
\node[circle,draw, inner sep=0 pt, minimum size=0.2cm](b1s2) at (-2.25,1) [label=left:$b_{1}^{s_{2}}$]{};

\node[circle,fill=orange,draw, inner sep=0 pt, minimum size=0.2cm](a3s2) at (0.75,2) [label=left:$a_{3}^{s_{2}}$]{};
\node[circle,fill=orange,draw, inner sep=0 pt, minimum size=0.2cm](a2s2) at (-0.75,2) [label=left:$a_{2}^{s_{2}}$]{};
\node[circle,fill=orange,draw, inner sep=0 pt, minimum size=0.2cm](a1s2) at (-2.25,2) [label=left:$a_{1}^{s_{2}}$]{};

\node[circle,fill=orange,draw, inner sep=0 pt, minimum size=0.2cm](b2s3) at (4.5,1) [label=left:$b_{2}^{s_{3}}$]{};
\node[circle,draw, inner sep=0 pt, minimum size=0.2cm](b1s3) at (3,1) [label=left:$b_{1}^{s_{3}}$]{};
\node[circle,fill=orange,draw, inner sep=0 pt, minimum size=0.2cm](b3s3) at (6,1) [label=left:$b_{3}^{s_{3}}$]{};

\node[circle,fill=white,draw, inner sep=0 pt, minimum size=0.2cm](a2s3) at (4.5,2) [label=left:$a_{2}^{s_{3}}$]{};
\node[circle,fill=white,draw, inner sep=0 pt, minimum size=0.2cm](a1s3) at (3,2) [label=left:$a_{1}^{s_{3}}$]{};
\node[circle,fill=white,draw, inner sep=0 pt, minimum size=0.2cm](a3s3) at (6,2) [label=left:$a_{3}^{s_{3}}$]{};

\node[circle,draw, inner sep=0 pt, minimum size=0.2cm](b1s1) at (-7.5,1) [label=left:$b_{1}^{s_{1}}$]{};
\node[circle,fill=orange,draw, inner sep=0 pt, minimum size=0.2cm](b2s1) at (-6,1) [label=left:$b_{2}^{s_{1}}$]{};
\node[circle,fill=orange,draw, inner sep=0 pt, minimum size=0.2cm](b3s1) at (-4.5,1) [label=left:$b_{3}^{s_{1}}$]{};

\node[circle,fill=white,draw, inner sep=0 pt, minimum size=0.2cm](a1s1) at (-7.5,2) [label=left:$a_{1}^{s_{1}}$]{};
\node[circle,fill=white,draw, inner sep=0 pt, minimum size=0.2cm](a2s1) at (-6,2) [label=left:$a_{2}^{s_{1}}$]{};
\node[circle,fill=white,draw, inner sep=0 pt, minimum size=0.2cm](a3s1) at (-4.5,2) [label=left:$a_{3}^{s_{1}}$]{};

\node[circle,fill=white, draw, inner sep=0 pt, minimum size=0.2cm](x) at (-1,4.5) [label=above:$x$]{};
\node[circle,fill=white, draw, inner sep=0 pt, minimum size=0.2cm](y) at (0,4.5) [label=above:$y$]{};

\node[circle, draw, inner sep=0 pt, minimum size=0.2cm](u1) at (-3.5,4) [label=above:$u_{1}$]{};

\node[circle, draw, inner sep=0 pt, minimum size=0.2cm](u2) at (1.5,4) [label=above:$u_{2}$]{};

\draw[green] (a1s1)--(a2s1);
\draw[green] (a3s1)--(a2s1);
\draw[green] (b1s1)--(b2s1);
\draw[green] (b3s1)--(b2s1);
\draw[green] (a1s1)--(b2s1);
\draw[green] (a1s1)--(b3s1);
\draw[green] (a2s1)--(b1s1);
\draw[green] (a2s1)--(b3s1);
\draw[green] (a3s1)--(b2s1);
\draw[green] (a3s1)--(b1s1);
\draw[green] (a1s1)--(b1s1);
\draw[green] (a2s1)--(b2s1);
\draw[green] (a3s1)--(b3s1);

\draw[green] (a1s2)--(a2s2);
\draw[green] (a3s2)--(a2s2);
\draw[green] (b1s2)--(b2s2);
\draw[green] (b3s2)--(b2s2);
\draw[green] (a1s2)--(b2s2);
\draw[green] (a1s2)--(b3s2);
\draw[green] (a2s2)--(b1s2);
\draw[green] (a2s2)--(b3s2);
\draw[green] (a3s2)--(b2s2);
\draw[green] (a3s2)--(b1s2);
\draw[green] (a1s2)--(b1s2);
\draw[green] (a2s2)--(b2s2);
\draw[green] (a3s2)--(b3s2);

\draw[green] (a1s3)--(a2s3);
\draw[green] (a3s3)--(a2s3);
\draw[green] (b1s3)--(b2s3);
\draw[green] (b3s3)--(b2s3);
\draw[green] (a1s3)--(b2s3);
\draw[green] (a1s3)--(b3s3);
\draw[green] (a2s3)--(b1s3);
\draw[green] (a2s3)--(b3s3);
\draw[green] (a3s3)--(b2s3);
\draw[green] (a3s3)--(b1s3);
\draw[green] (a1s3)--(b1s3);
\draw[green] (a2s3)--(b2s3);
\draw[green] (a3s3)--(b3s3);

\draw[green](a1s1)..controls(-6,2.8)..(a3s1);
\draw[green](b1s1)..controls(-6,0.2)..(b3s1);

\draw[green](a1s2)..controls(-.75,2.8)..(a3s2);
\draw[green](b1s2)..controls(-.75,0.2)..(b3s2);

\draw[green](a1s3)..controls(4.5,2.8)..(a3s3);
\draw[green](b1s3)..controls(4.5,0.2)..(b3s3);

\draw(u1)--(a2s1);
\draw(u1)--(a3s1);
\draw(u1)--(a1s2);
\draw(u1)--(a1s3);

\draw(u2)--(a3s1);
\draw(u2)--(a2s2);
\draw(u2)--(a2s3);
\draw(u2)--(a3s3);

\draw(u1)--(x);
\draw(u2)--(x);

\draw(x)--(y);

\end{tikzpicture}
    \caption{The graph $G$ in the proof of Theorem \ref{trtheorem-cvd} constructed from MRSS instance $S=\{ (2,1),(1,1),(1,2)\}, g=(3,3), k=2, k'=2$. The set $S'=\{(2,1),(1,2)\}$ forms a solution of MRSS instance and the set
    $H= B^{s_1}\setminus b_{1}^{s_{1}} \cup B^{s_3}\setminus b_{1}^{s_{3}} \cup A^{s_2} $ forms a harmless set in $G$. }
    \label{thorem7}
\end{figure}

For each vector
$s\in S$, we introduce a clique $K^s$. 
For $s\in S$, $ \max(s)$ is the value of the largest coordinate of $s$ and
$\max(S)$ is maximum of $\max(s)$ values. The clique $K^s$ consists of vertices $V(K^s) = A^s \cup B^s$ where $A^s=\{a^s_{1},\ldots,a^s_{\max(S)+1}\}$ and $B^s=\{b^s_{1},\ldots,b^s_{\max(S)+1}\}$. 
 Next, we introduce the set $U=\{u_{1},\ldots,u_{k}\}$ of vertices into $G$. 
  For each $1\leq i \leq k$ and $s\in S$, 
  make $u_i$ adjacent to exactly $s(i)$ many vertices of $A^s$ arbitrarily. 
Finally, we add two vertices $x$ and $y$. Next, make $x$ adjacent to all the vertices in set $U$ and $y$. This completes the construction of graph $G$. 
Note that deletion of the set $U$ of size $k$ from $G$ results in a disjoint union of complete graphs.
  We define the threshold function as follows:

 \[ t(u) = 
            \begin{cases}
                1 &\quad\text{if $u \in \{x,y\}$}\\
               \max(S)+1 &\quad\text{if $u \in \bigcup\limits_{s\in S} A^s$}\\
               \max(S)+2 & \quad\text{if $u\in \bigcup\limits_{s\in S} B^s $ }\\
               \sum\limits_{s\in S} s(i) - g(i)+1 & \quad\text{if $u= u_{i}$ for all $1\leq i \leq k$}
            \end{cases} 
            \]
We set $r= k'\times \max(S)+ (n-k')\times (\max(S)+1)$. Now we show that our reduction is correct. That is, we claim $(k, k', S, g)$ is  a yes instance of 
{\sc MRSS} if and only if $(G,t,r)$ is a yes instance of {\sc Harmless Set}. 
Towards showing the forward direction, let $S'\subseteq S$ be such that 
$|S'|\leq k'$ and $\sum\limits_{s\in S'}{s}\geq g$. 
We claim that 
$$H = \bigcup\limits_{s\in S'} B^s\backslash \{b_{1}^{s}\} \bigcup\limits_{s\in S\setminus S'} A^s$$ is a harmless set in $G$ of size at least $r$.
Clearly $|H|\geq r$. Next, we show that all vertices of $G$ satisfy 
the threshold condition. Since $x$ and $y$ have no neighbours in $H$, 
they satisfy the threshold condition. If $s\in S'$,
then every vertex of $A^{s} \cup B^{s}$ has at most $\max(S)$ neighbours in $H$.
Therefore, they satisfy the threshold condition. 
If $s\in S\setminus S'$, 
then every vertex of $A^{s}$ has exactly $\max(S)$ neighbours in $H$ and every vertex of $B^{s}$ has exactly $\max(S)+1$ neighbours in $H$.
Therefore, all vertices in cliques $K^{s}$ satisfy the threshold condition. 
Let $u_{i}$ be an arbitrary  element in  $U$. 
As $\sum\limits_{s\in S'}{s(i)}\geq g(i)$, we get 
$d_H(u_i)= \sum\limits_{s\in S \setminus S'}{s(i)}= \sum\limits_{s\in S }{s(i)}- \sum\limits_{s\in S'} {s(i)} \leq \sum\limits_{s\in S}s(i)-g(i)<t(u_i)$.

\par Towards showing the reverse direction, let $H$ be a harmless 
set of size at least $r$ in $G$. It is easy to see that 
$H \cap (U \cup \{x,y\}) = \emptyset$ as otherwise one of the 
vertices in $ \{x,y\}$  will fail to satisfy the threshold condition. Observe that any clique $K^{s}$ can contribute at most $\max(S)+1$ vertices as otherwise vertices in set $A^{s}$ will fail to satisfy the threshold condition. We prove the following simple claim.

\begin{claim}
If $|K^{s}\cap H|=\max(S)+1$ then $K^{s}\cap H = A^{s}$.
\end{claim}

\proof  Targeting a contradiction, assume that there exists a vertex 
$a^{s}\in A^{s}$ such that $a^{s} \not\in H$. 
As $|K^{s}\cap H|=\max(S)+1$, we have
 $d_{H}(a^{s}) =\max(S)+1 = t(a^{s})$, which is a contradiction.

\noindent 

\noindent Note that $n$ cliques together contributes at least $r= k'\times \max(S)+ (n-k')\times (\max(S)+1)$ vertices to $H$ and each clique can contributes at most $\max(S)+1$ vertices.
Therefore, by Pigeonhole principle, there are at least $(n-k')$ cliques $K^{s}$ such that  $|H \cap K^{s}| = \max(S)+1$. By the above claim, there are at least $(n-k')$ cliques $K^{s}$
such that $H \cap K^{s} = A^{s}$.
 We define
$$S'=\{s\in S~|~ H \cap K^{s} \neq A^{s}\}.$$
Clearly for  $s\in S'$, we have $|K^{s} \cap H|\leq \max(S)$. It is easy to see  that  $$H'=\big(H \setminus \bigcup\limits_{s\in S'} A^{s}\big) \cup \bigcup\limits_{s\in S'} B^{s}\setminus \{b^s_1\}$$ is again a harmless set with $|H'|\geq |H|$. 
Every vertex of $K^{s}$, $s\in S'$, satisfies the threshold condition because $|K^{s} \cap H'|=\max(S)$. For each $u_i\in U$, we see that $d_{H'}(u_{i}) \leq d_{H}(u_{i}) < t(u_{i})$. For rest of the vertices, we can easily verify that the threshold conditions are satisfied. This implies that $H'$ is a harmless set of size at least $r$. 
So we consider the harmless set to be of the form
$H'=  \bigcup\limits_{s\in S\setminus S'} A^s \bigcup\limits_{s\in S'} B^{s}\setminus \{b_{1}^{s}\}$. Since  each $u_i\in U$ satisfies the 
threshold condition, we have $d_{H'}(u_i)= \sum\limits_{s\in S\setminus S'}{s(i)}=
\sum\limits_{s\in S}{s(i)} -\sum\limits_{s\in S'}{s(i)}<t(u_i)=
\sum\limits_{s\in S} s(i)-g(i)+1$. This implies that $\sum\limits_{s\in S'}s(i)\geq g(i)$ for  $1\leq i\leq k$. Therefore  $(k, k', S, g)$ is a yes-instance. \qed \\
\section{Graphs of bounded  clique-width} 
This section presents a polynomial time 
algorithm for the {\sc Harmless set} 
problem  for graphs of bounded clique-width. 
The clique-width of a graph $G$, denoted by ${\tt cw}(G)$, is the minimum number of labels needed 
to construct $G$ 
using the following four operations:
\begin{enumerate}
    \item Create a new graph with a single vertex $v$ with label $i$ (written $i(v)$).
    \item Take the disjoint union of two labelled graphs $G_1$ and $G_2$ (written $G_1 \cup G_2$).
    \item Add an edge between every vertex with label $i$ and every vertex with label $j$,
$i\neq j$ (written $\eta_{ij}$).
\item Relabel every vertex with label $i$ to have label $j$ (written $\rho_{i\rightarrow j}$).
\end{enumerate}
We say that a construction of a graph $G$ with the four operations is a $c$-expression if it 
uses at most $c$ labels. Thus the clique-width of $G$ is the minimum $c$ for which $G$ has a 
$c$-expression.  
A $c$-expression is a rooted binary tree $T$ such that
\begin{enumerate}
    \item each leaf has label $i$ for some $i\in \{1,\ldots,c\}$,
    \item each non-leaf node with two children has label $\cup $, and 
    \item each non-leaf node with only one child has label $\rho_{i\rightarrow j}$ 
    or $\eta_{i,j}$ $(i,j\in \{1,\ldots,c\}, i\neq j)$.
\end{enumerate}
\begin{example} Consider the graph $P_n$, which is simply a path on $n$ vertices. 
Note that ${\tt cw}(P_1) = 1$ and ${\tt cw}(P_2) = {\tt cw}(P_3) = 2$. 
Now consider a path on four vertices $v_1, v_2, v_3, v_4$, in that order. Then this path can be constructed using 
the four operations (using only three labels) as follows:
$$\eta_{3,2}(3(v_4) \cup \rho_{3\rightarrow 2}(\rho_{2\rightarrow 1}(\eta_{3,2}(3(v_3) \cup \eta_{2,1}(2(v_2) \cup  1(v_1)))))).$$
This construction can readily be generalized to longer paths for $n \geq  5$. It is easy to see that 
$cw(P_n)=3$ for all $n\geq 4$.
\end{example} 
A $c$-expression represents the graph represented by its root. A $c$-expression 
of a $n$-vertex graph $G$ has $O(n)$ vertices. 
A $c$-expression of a graph is {\it  irredundant}  if for each edge $\{u,v\}$, there is exactly 
one node $\eta_{i,j}$ that adds the edge between $u$ and $v$. It is known that a $c$-expression 
of a graph can be transformed into an irredundant one with $O(n)$ nodes in linear time. 
Here we use irredundant $c$-expression only. 

Computing the clique-width and a corresponding $c$-expression of a graph is NP-hard. 
For $c\leq 3$, we can compute a $c$-expression  of a graph of clique-width at most $c$
in $O(n^2m)$ time, where $n$ and $m$ are the number of vertices and edges, respectively. 
For fixed $c\geq 4$, it is not known whether one can compute the clique-width and a corresponding 
$c$-expression of a graph in polynomial time. On the 
other hand, it is known that for any fixed $c$, one can compute a 
$(2^{c+1}-1)$-expression of a graph of clique-width $c$
in $O(n^3)$ time. 
For more details see \cite{KAMINSKI20092747}.

\begin{theorem}\label{cliqueth}
Given an $n$-vertex graph $G$ and an irredundant $c$-expression $T$ of $G$, the {\sc Harmless Set} problem is solvable in $O(n^{4c})$ time. 
\end{theorem}

For each node $t$ in a $c$-expression $T$, let $G_t$ be the vertex-labeled graph represented by $t$. 
We denote by $V_t$ the vertex set of $G_t$. For each $i$, we denote the set of 
$i$-vertices in $G_t$ by $V_t^i$. For each node $t$ in $T$, we construct a table $dp_t({\tt r, s}) \in \{\mbox{true, false}\} $
with indices ${\tt r}: \{1,\ldots,c\}\rightarrow \{0,\ldots,n\}$ and
 ${\tt s}: \{1,\ldots,c\}\rightarrow \{-n+1,\ldots,n-1\}\cup \{\infty\}$
as follows. We set $dp_t({\tt r, s})=\mbox{true} $ if and only if there exists a set  $S$ in $ V_t$ such that 
for all $i\in\{1,2,\ldots,c\}$
\begin{itemize}
    \item ${\tt r}(i)=|S\cap V_t^i|$;
    \item ${\tt {s}}(i)=\mbox{min}_{v\in V_t^i}\Big\{  t(v)-|N_{G_t}(v)\cap S|\Big\}$, otherwise 
    ${\tt {s}}(i)=\infty$.    
\end{itemize}
That is, ${\tt r}(i)$ denotes the number of the $i$-vertices in $S$ and  ${\tt {s}}(i)$ is the ``surplus" at the weakest $i$-vertex in $S$. 
\par Let $\tau$ be the root of the $c$-expression $T$ of $G$. Then  $G$ has a harmless set of size $h$
if there exist ${\tt r, s} $ satisfying 
\begin{enumerate}
    \item $dp_{\tau}({\tt r, s})=\mbox{true} $;
    \item $\sum_{i=1}^{c} {\tt} r(i)=h $
    \item $\mbox{min}\Big\{ {\tt s}(i)\Big\}\geq 1$. 
\end{enumerate} 

In the following, we compute all entries $dp_{t}({\tt r, s})$
in a bottom-up manner. There are $(n+1)^c \cdot (2n)^c=$ $O(n^{2c})$ 
possible tuples $({\tt r, s})$. Thus, to prove Theorem \ref{cliqueth}, it is enough to prove that 
each entry $dp_{t}({\tt r,s})$ can be computed in time $O(n^{2c})$ assuming that 
the entries for the children of $t$ are already computed.

\begin{lemma}\rm 
For a leaf node $t$ with label $i$, $dp_t({\tt r, s})$ can be computed in $O(1)$ time. 
\end{lemma}
\proof Observe that $dp_t({\tt r, s})={\tt true}$ if and only if ${\tt r}(j)=0$, ${\tt s}(j)=\infty$ for all $j\neq i$, and either
  \begin{itemize}  
  \item   ${\tt r}(i)=0$, ${\tt s}(i)=\infty$ or 
 \item  ${\tt r}(i)=1$, ${\tt s}(i)\geq 1$.  
 \end{itemize}
The first case corresponds to $S=\emptyset$, and the second case 
corresponds to $S=V_t^i$. These conditions can be checked in $O(1)$ time. 

\begin{lemma}\rm 
For a $\cup$ node $t$, $dp_t({\tt r, s})$ can be computed in $O(n^{2c})$ time.
\end{lemma}

\proof Let $t_1$ and $t_2$ be the children of $t$ in $T$. Then $dp_t({\tt r, s})={\tt true}$ 
if and only if there exist ${\tt r_1, s_1}$ and ${\tt r_2, s_2}$ such that 
$dp_t({\tt r_1, s_1})={\tt true}$,  $dp_t({\tt r_2, s_2})={\tt true}$,
${\tt r}(i)={\tt r_1}(i)+{\tt r_2}(i)$,
${\tt s}(i)=\mbox{min}\Big\{ {\tt s}_1(i), {\tt s}_2(i)\Big\}$ for all $i$. The number of possible pairs for 
${\tt (r_1,r_2)}$ is  at most $(n+1)^c$ as ${\tt r_2}$ is uniquely determined by
${\tt r_1}$.  There are at most $(2n)^c$ possible pairs for $({\tt s_1,s_2})$.  In total, there are $O(n^{2c})$ candidates. Each candidate can be checked in 
$O(1)$ time, thus the lemma holds. 

\begin{lemma}\rm 
For a $\eta_{ij}$ node $t$, $dp_t({\tt r, s})$ can be computed in $O(1)$ time.
\end{lemma}

\proof Let $t^{\prime}$ be the child of $t$ in $T$. Then,  
$dp_t({\tt r,s})={\tt true}$ if and only if \\
$dp_t({\tt r, s^{\prime}})={\tt true}$ 
for some ${\tt s^{\prime}}$ with the following conditions:
\begin{itemize}
    \item ${\tt s}(h)={\tt s}^{\prime}(h)$ hold for all $h\notin \{i,j\}$;
    \item ${\tt s}(i)={\tt s^{\prime}}(i)-{\tt r}(j)$ and  ${\tt s}(j)={\tt s^{\prime}}(j)-{\tt r}(i)$. 
    \end{itemize}
 We now explain the condition for $s(i)$. Recall that $T$ is irredundant. That is, the graph $G_{t^{\prime}}$ does not have 
any edge between the $i$-vertices and the $j$-vertices. In $G_t$, an $i$-vertex has exactly
${\tt r}(j)$ more neighbours in $S$ and similarly a $j$-vertex has exactly
${\tt r}(i)$ more neighbours in $S$.
Thus we have ${\tt s}(i)={\tt s^{\prime}}(i)-{\tt r}(j)$ and 
${\tt s}(j)={\tt s^{\prime}}(j)-{\tt r}(i)$. The lemma holds as there is only one candidate for each 
${\tt s}^{\prime}(i)$ and ${\tt s}^{\prime}(j)$.

\begin{lemma}\rm 
For a $\rho_{i\rightarrow j}$ node $t$, $dp_t({\tt r, s})$ can be computed in $O(n^2)$ time.
\end{lemma}

\proof Let $t^{\prime}$ be the child of $t$ in $T$. Then,  $dp_t({\tt r, s})={\tt true}$ 
if and only if there exist  ${\tt r}^{\prime}, {\tt s}^{\prime}$ such that 
$dp_{t^{\prime}}({\tt r^{\prime}}, {\tt s}^{\prime})
={\tt true}$, where :
\begin{itemize}
\item ${\tt r}(i)=0 $, ${\tt r}(j)={\tt r}^{\prime}(i)+{\tt r}^{\prime}(j)$, and 
${\tt r}(h)={\tt r}^{\prime}(h)$ if $h\notin \{i,j\}$;
\item ${\tt s}(i)=\infty $, ${\tt s}(j)=\mbox{min}\big\{{\tt s}^{\prime}(i), {\tt s}^{\prime}(j)\big\}$, and 
${\tt s}(h)={\tt s}^{\prime}(h)$ if $h\notin \{i,j\}$.
\end{itemize}
The number of possible pairs for $({\tt r}^{\prime}(i), {\tt r}^{\prime}(j))$ is $O(n)$ as 
${\tt r}^{\prime}(j)$ is uniquely determined by ${\tt r}^{\prime}(i)$.
There are at most $O(n)$ possible pairs for $({\tt s^{\prime}}(i),{\tt s^{\prime}}(j))$.  In total, there are $O(n^2)$ candidates. Each candidate can be checked in 
$O(1)$ time, thus the lemma holds. \\

\section{{\sc Harmless Set} on Planar Graphs}

In this section, we propose a fixed parameter tractable algorithm for {\sc Harmless Set} parameterized by the solution size, even when restricted to planar graphs. Note that the {\sc Harmless Set} problem parameterized by the solution size is W[1]-hard on general graphs even when all thresholds are bounded by a constant.

\begin{theorem}
The {\sc Harmless Set} problem with general thresholds
parameterized by solution size  is fixed parameter tractable on planar graphs.
\end{theorem}

\proof Let $(G,k)$ be an instance of {\sc Harmless Set}, where $G$ is a planar 
graph. First we do some preprocessing on  $G$. If  $v\in V(G)$ 
has a neighbour with threshold value 1 then clearly $v$ cannot be part of any harmless set; we color $v$ red.  We color the rest of the vertices green. 
We observe that if a vertex and all of its neighbours are colored red then 
its removal does not change the solution. This shows that the following rule is safe.

\begin{itemize}
    \item Reduction 1: If a vertex $v$ and  its neighbours are colored red then delete $v$ from $G$. The new  instance is $(G-v,k)$.
\end{itemize}
Next we claim that after exhaustive application of Reduction  1 if the diameter of the reduced graph $G$ is greater than or equal to $6k$ then we always get a yes instance. Let us assume that the diameter of graph $G$ is   $6k$. 
Then there exists a pair of nonadjacent vertices $a$ and $b$ such that $d(a,b)=6k$. Let $P=(v_{1},v_{2},\ldots,v_{6k+1})$ be a shortest 
path joining $a$ and $b$, where $v_{1}=a$ and $v_{6k+1}=b$. Now, we  construct a harmless set $S$ of size $k$ containing only green vertices. 
Since we cannot apply Reduction 1, every vertex on $P$  is either green or at least one of its neighbours is green. 
For every vertex $v_{6i+1} \in P$, $0\leq i\leq k$, 
if $v_{6i+1}$ is colored green we include it in $S$; otherwise  
if $v_{6i+1}$ is colored  red, we include one of its green neighbours in $S$. 
We claim that $S$ is a harmless set of size $k$. Clearly, $|S|\geq k$. 
Since we include only green vertices in $S$, all the vertices with threshold one satisfy the threshold condition. Next, consider a vertex $u$ with threshold at least two. We show that $d_S(u)\leq 1$. Assume, for the sake of contradiction, that $u$ is adjacent to two vertices $s_{1},s_{2}\in S$. We can assume that $s_{1}\in N_{G}[v_{6i+1}]$ and 
$s_{2}\in N_{G}[v_{5(i+i)+1}]$ for some $i$. Note that, we have $d(v_{6i+1},v_{6(i+1)+1})=5$ by construction. If $u$ is adjacent to $s_{1}$ and $s_{2}$ then it implies that $d(v_{6i+1},v_{6(i+1)+1})\leq 4$ and hence $d(a,b)<6k$, which is a contradiction. This shows that all the vertices with threshold greater than or equal to two also satisfy the threshold condition. Therefore, we get a harmless set $S$ of size at least $k$. Based on the above argument, our second rule is the following.
\begin{itemize}
    \item Reduction 2: If the diameter of  $G$ is more than or equal to $6k$ then
    we conclude that we are dealing with a  yes-instance.
\end{itemize}
Let $(G,k)$ be an input instance such that Reduction 1 and 2 are not applicable to 
$(G,k)$. Then $G$ is a planar graph with diameter  at most $6k$, it implies that treewidth of $G$ is at most $18k$. Now,  we can solve the problem using a standard dynamic programming algorithm technique in  time $2^{\mathcal{O}(k \log{k})}$. 
Note that when we are looking for a harmless set of size $k$, we can assume that $t_{\max}\leq k+1$ where $t_{\max}$ denotes the value of maximum threshold. 

\section{Conclusion and Future Directions}
We have shown that  {\sc Harmless Set} with general thresholds is W[1]-hard
parameterized by the size of a vertex deletion set into
trees of height at most 3
 and also the cluster vertex deletion number of the input graph. On the positive side, we have given FPT algorithms when parameterized by any of the following parameters: vertex integrity, neighbourhood diversity and twin cover. To give an upper bound on the complexity, we give an XP-algorithm when parameterized by cliquewidth. For the {\sc Harmless Set} problem with majority thresholds, we have  shown W[1]-hardness parameterized by treewidth. 
\par The natural next step is to figure out the parameterized complexity of {\sc Harmless Set} for general thresholds with respect to the parameters such as vertex deletion to disjoint paths, modular width and co-cluster vertex deletion set. For {\sc Harmless Set}  with majority thresholds, it will be interesting to see if the parameters such as treedepth, feedback vertex set and cluster vertex deletion set allow FPT algorithms or the problem still remains hard.

\bibliographystyle{abbrv}
\bibliography{bibliography}

\begin{thebibliography}{10}

\bibitem{Aazami}
A.~Aazami and K.~Stilp.
\newblock Approximation algorithms and hardness for domination with
  propagation.
\newblock {\em SIAM Journal on Discrete Mathematics}, 23(3):1382--1399, 2009.

\bibitem{Bazgan-DO}
C.~Bazgan and M.~Chopin.
\newblock The complexity of finding harmless individuals in social networks.
\newblock {\em Discret. Optim.}, 14(C):170–182, Nov. 2014.

\bibitem{BazganCocoon}
C.~Bazgan, M.~Chopin, A.~Nichterlein, and F.~Sikora.
\newblock Parameterized approximability of maximizing the spread of influence
  in networks.
\newblock In D.-Z. Du and G.~Zhang, editors, {\em Computing and Combinatorics},
  pages 543--554, Berlin, Heidelberg, 2013. Springer Berlin Heidelberg.

\bibitem{BENZWI201187}
O.~Ben-Zwi, D.~Hermelin, D.~Lokshtanov, and I.~Newman.
\newblock Treewidth governs the complexity of target set selection.
\newblock {\em Discrete Optimization}, 8(1):87--96, 2011.
\newblock Parameterized Complexity of Discrete Optimization.

\bibitem{CENTENO20113693}
C.~C. Centeno, M.~C. Dourado, L.~D. Penso, D.~Rautenbach, and J.~L.
  Szwarcfiter.
\newblock Irreversible conversion of graphs.
\newblock {\em Theoretical Computer Science}, 412(29):3693--3700, 2011.

\bibitem{ChenSIAM}
N.~Chen.
\newblock On the approximability of influence in social networks.
\newblock {\em SIAM Journal on Discrete Mathematics}, 23(3):1400--1415, 2009.

\bibitem{Chiang}
C.-Y. Chiang, L.-H. Huang, B.-J. Li, J.~Wu, and H.-G. Yeh.
\newblock Some results on the target set selection problem.
\newblock {\em Journal of Combinatorial Optimization}, 25(4):702--715, 2013.

\bibitem{ChopinTCS}
M.~Chopin, A.~Nichterlein, R.~Niedermeier, and M.~Weller.
\newblock Constant thresholds can make target set selection tractable.
\newblock {\em Theory of Computing Systems}, 55(1):61--83, 2014.

\bibitem{marekcygan}
M.~Cygan, F.~V. Fomin, L.~Kowalik, D.~Lokshtanov, D.~Marx, M.~Pilipczuk,
  M.~Pilipczuk, and S.~Saurabh.
\newblock {\em Parameterized Algorithms}.
\newblock Springer, 2015.

\bibitem{Downey}
R.~G. Downey and M.~R. Fellows.
\newblock {\em Parameterized Complexity}.
\newblock Springer, 2012.

\bibitem{Pal}
P.~G. Drange, M.~Dregi, and P.~van~'t Hof.
\newblock On the computational complexity of vertex integrity and component
  order connectivity.
\newblock {\em Algorithmica}, 76(4):1181--1202, 2016.

\bibitem{DREYER20091615}
P.~A. Dreyer and F.~S. Roberts.
\newblock Irreversible k-threshold processes: Graph-theoretical threshold
  models of the spread of disease and of opinion.
\newblock {\em Discrete Applied Mathematics}, 157(7):1615--1627, 2009.

\bibitem{fellows}
M.~R. Fellows, D.~Lokshtanov, N.~Misra, F.~A. Rosamond, and S.~Saurabh.
\newblock Graph layout problems parameterized by vertex cover.
\newblock In S.-H. Hong, H.~Nagamochi, and T.~Fukunaga, editors, {\em
  Algorithms and Computation}, pages 294--305, Berlin, Heidelberg, 2008.
  Springer Berlin Heidelberg.

\bibitem{mss}
R.~Ganian, F.~Klute, and S.~Ordyniak.
\newblock On structural parameterizations of the bounded-degree vertex deletion
  problem.
\newblock {\em Algorithmica}, 2020.

\bibitem{Gima}
T.~Gima, T.~Hanaka, M.~Kiyomi, Y.~Kobayashi, and Y.~Otachi.
\newblock Exploring the gap between treedepth and vertex cover through
  vertex integrity.
\newblock In T.~Calamoneri and F.~Cor{\`o}, editors, {\em Algorithms and
  Complexity}, pages 271--285, Cham, 2021. Springer International Publishing.

\bibitem{KAMINSKI20092747}
M.~Kami{\'n}ski, V.~V. Lozin, and M.~Milani{\v c}.
\newblock Recent developments on graphs of bounded clique-width.
\newblock {\em Discrete Applied Mathematics}, 157(12):2747 -- 2761, 2009.

\bibitem{kannan}
R.~Kannan.
\newblock Minkowski's convex body theorem and integer programming.
\newblock {\em Mathematics of Operations Research}, 12(3):415--440, 1987.

\bibitem{KempeTardos}
D.~Kempe, J.~Kleinberg, and E.~Tardos.
\newblock Maximizing the spread of influence through a social network.
\newblock In {\em Proceedings of the Ninth ACM SIGKDD International Conference
  on Knowledge Discovery and Data Mining}, KDD '03, page 137–146, New York,
  NY, USA, 2003. Association for Computing Machinery.

\bibitem{Kloks94}
T.~Kloks.
\newblock {\em Treewidth, Computations and Approximations}, volume 842 of {\em
  Lecture Notes in Computer Science}.
\newblock Springer, 1994.

\bibitem{Lampis}
M.~Lampis.
\newblock Algorithmic meta-theorems for restrictions of treewidth.
\newblock {\em Algorithmica}, 64:19--37, 2012.

\bibitem{lenstra}
H.~W. Lenstra.
\newblock Integer programming with a fixed number of variables.
\newblock {\em Mathematics of Operations Research}, 8(4):538--548, 1983.

\bibitem{Sparsity}
J.~Nesetril and P.~O. de~Mendez.
\newblock {\em Sparsity: Graphs, Structures, and Algorithms}.
\newblock Springer Publishing Company, Incorporated, 2014.

\bibitem{Nichterlein}
A.~Nichterlein, R.~Niedermeier, J.~Uhlmann, and M.~Weller.
\newblock On tractable cases of target set selection.
\newblock {\em Social Network Analysis and Mining}, 3(2):233--256, 2013.

\bibitem{PELEG}
D.~Peleg.
\newblock Local majorities, coalitions and monopolies in graphs: a review.
\newblock {\em Theoretical Computer Science}, 282(2):231 -- 257, 2002.

\bibitem{JGAA-244}
T.~{Reddy} and C.~{Rangan}.
\newblock Variants of spreading messages.
\newblock {\em Journal of Graph Algorithms and Applications}, 15(5):683--699,
  2011.

\bibitem{Neil}
N.~Robertson and P.~Seymour.
\newblock Graph minors. iii. planar tree-width.
\newblock {\em Journal of Combinatorial Theory, Series B}, 36(1):49 -- 64,
  1984.

\bibitem{DBLP:journals/corr/abs-1107-1177}
S.~Szeider.
\newblock Not so easy problems for tree decomposable graphs.
\newblock {\em CoRR}, abs/1107.1177, 2011.

\bibitem{Tedder}
M.~Tedder, D.~Corneil, M.~Habib, and C.~Paul.
\newblock Simpler linear-time modular decomposition via recursive factorizing
  permutations.
\newblock In {\em Automata, Languages and Programming}, pages 634--645, Berlin,
  Heidelberg, 2008. Springer Berlin Heidelberg.

\end{thebibliography}
\end{document}